\documentclass[aps,prb,reprint,showpacs,superscriptaddress,groupedaddress]{revtex4-1}
\usepackage{graphicx}
\usepackage{amsmath}
\usepackage{amssymb}
\usepackage{dcolumn}
\usepackage{latexsym}
\usepackage{rotating}
\usepackage[usenames,dvipsnames]{xcolor}
\usepackage{float}
\usepackage{epsfig}
\usepackage{psfrag}
\usepackage{natbib}
\usepackage{bm}
\usepackage{eucal}
\usepackage{braket}
\usepackage{enumerate}
\usepackage{longtable}
\usepackage{subfigure}
\usepackage{bm}
\usepackage{hyperref}
\usepackage{amsfonts}
\usepackage{dcolumn}
\usepackage{bm}
\usepackage{subfigure}

\newcommand{\be}{\begin{equation}}
\newcommand{\ee}{\end{equation}}
\newcommand{\bn}{\begin{eqnarray}}
\newcommand{\en}{\end{eqnarray}}

\usepackage{color} 

\usepackage{hyperref}

\begin{document}
\author{S. Koley$^{1}$}

\title{Engineering Si doping in anatase and rutile $TiO_2$ with oxygen vacancy 
for efficient optical application}
\affiliation{$^{1}$ Department of Physics, North Eastern Hill University,
Shillong, Meghalaya, 793022 India}

\begin{abstract}
\noindent In this paper, I demonstrate a density functional theory plus 
dynamical mean field theory
study on the electronic properties of doped $TiO_2$ rutile as well 
as another tetragonal phase anatase with oxygen vacancy. The density of states and optical
 properties have been obtained from the electronic structure applying 
screened hybrid exchange correlation density functionals. All the
single-particle excitations are treated within the dynamical mean field theory 
for independent quasiparticles. 
For optical properties, excitations are considered by solving the 
Bethe-Salpeter equation for Coulomb correlated electron-hole duo. On this 
theoretical basis, band structure and optical spectra for the two structures of 
$TiO_2$ are provided. Further, I compared the present results with earlier 
optical data of parent structure and established the increased optical 
efficiency in doped $TiO_2$ with oxygen vacancy in both the structure.
\end{abstract}
\maketitle

\noindent Nanoparticles exhibit surprising electronic and magnetic properties 
and thus they are having a number of applications in the industrial sectors like
 opto-electronics, bio-medical, cosmetics and many others\cite{nano}. The properties like 
hardness, rigidity, surface to volume ratio distinguish nanoparticles from its 
bulk systems. $TiO_2$ is one of the most renowned materials among the oxides for 
its variety of industrial applications. The transition metal-oxygen bond here 
is having a property of increasing co-valency with oxygen. Its catalytic 
property is used in photocatalysis for organic synthesis. Moreover $TiO_2$ is 
largely 
used in white painting, sensors, photovoltaic devices, food preservatives or 
coloring, cosmetics and in cancer treatment\cite{hoff,maldoti,bonhole,phillips}.

\noindent The $TiO_2$ polymorphs are predominantly three crystal types, 
rutile (tetragonal ($4/mmm$), space group: $P4_2/mnm–D^{14}_{4h}$ ), brookite
(orthorhombic ($mmm$), $Pbca–D^{15}_{2h}$)
and anatase (tetragonal ($4/mmm$), $I4_1/amd–D^{19}_{4h}$). Among them only
rutile and anatase have a major role in industry based applications.
Experimental results on $TiO_2$ brookite is little short due to its
tough preparations and rare appearance\cite{chemmater}.
The nanostructure of this transition metal oxide material affects the 
phase behavior and the thermal stability of the material. Moreover it also 
has influence in industrial applications. Nanostructure of 
this TMO shows band gap changes as inverse square of particle size\cite{luca}.
The electrical conductivity is also affected due to the grain boundaries.
Since it is a non-toxic semiconductor material with cost effective availability,
 and long term stability $TiO_2$ has been taken into account for
photovoltaic applications. The optical gaps are reported to be 
slightly above 3 eV for all the three polymorphs (rutile: 3.0 eV, anatase:
3.4 eV and brookite: 3.3 eV),\cite{3,5,6} so $TiO_2$ shows
photoactivity in the UV region and acts as inefficient active solar cell. 
Though in dye-sensitized solar cells it acts as an economically viable and good 
photoactive dye\cite{7,8,9}.

\noindent Since $TiO_2$ is an important semiconductor having industrial 
applications with large band-gap and frequently used in daily life, 
enormous research has been already reported\cite{tio2review}. Also $TiO_2$ has 
good applications 
for photocatalytic reactions, including earth abundance and resistance to 
photocorrosion\cite{chen}. But there is a great disadvantage of using $TiO_2$ owing to its 
 low quantum effficiency in photocatalytic reactions and high recombination of 
photogenerated electron-hole pairs. 
So a lot of methods have been used to modify the properties of
$TiO_2$ so that the lifetime of photogenerated electron-hole pairs increase 
and reduce the band gap. Among them doping $TiO_2$\cite{60,64}, co-doping with two or 
more ions\cite{chen,65},  hybridization with carbon materials\cite{66,70,72}, coupling with small band gap 
semiconductor\cite{25,29} are known to make metal core $TiO_2$ shell
composite photocatalysts. Recently defect induced changes in light absorption properties are also reported\cite{77,79}.

Among all other defects found in $TiO_2$ , oxygen vacancies are very important 
and appear in many metal oxides\cite{80}. Oxygen vacancies are easy to manipulate 
in metal oxides and can change the properties of oxides drastically. Both
 theoretical calculations and experimental characterizations \cite{81,83,84} 
prove that oxygen
vacancies in $TiO_2$ are essential for photocatalysis and can lead to 
ferromagnetism and act as a magnetic semiconductor in spintronics. In 
principle, the oxygen vacancies in $TiO_2$ lead to the formation of unpaired
electrons or $Ti^{3+}$ centers, which also make donor levels in the
band structure of $TiO_2$. Additionally, oxygen vacancies are
believed to affect the electron-hole recombination process in
photocatalysts, causing a change in chemical rates that
depends on charge transfer from either electrons or holes. 
Several investigations also reported that oxygen vacancy result in 
metallic conductivity \cite{gu} at room temerature but resistivity 
increases with cooling.

For all the above reasons, still there is a growing interest in the
development of oxygen vacancy and doping in $TiO_2$ and thereby exploring the 
properties 
for orbital composition and magnetic moment.
In this article I will focus on these for the defect induced $TiO_2$ polymorphs 
rutile and anatase from theory using Si as dopant. I will employ a combination 
of density functional theory(DFT) and 
dynamical mean field theory(DMFT) and find out optical constants to show increased efficiency. 

\begin{figure}
(a)
\includegraphics[angle=270,width=0.9\columnwidth]{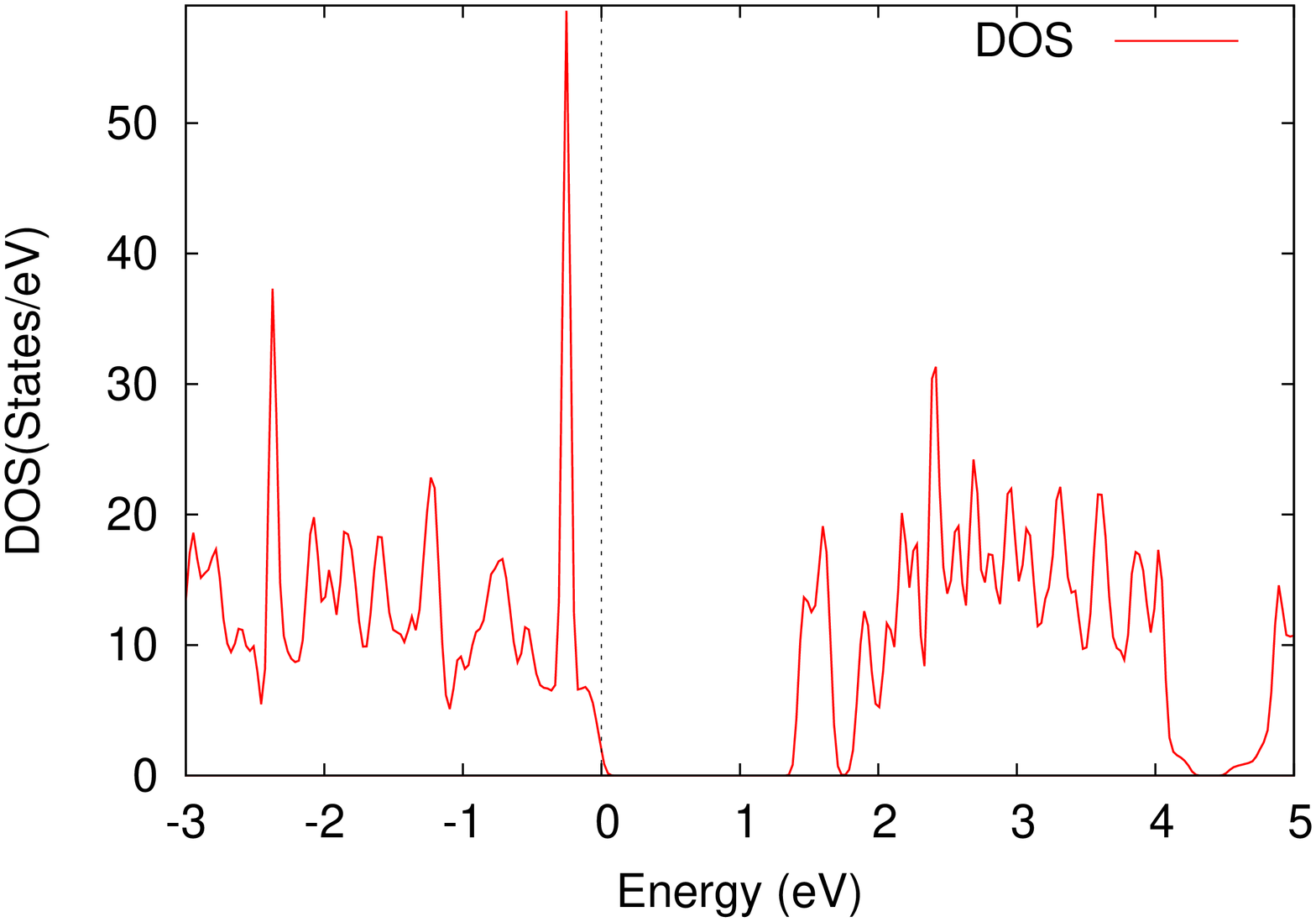}
(b)
\includegraphics[angle=270,width=0.9\columnwidth]{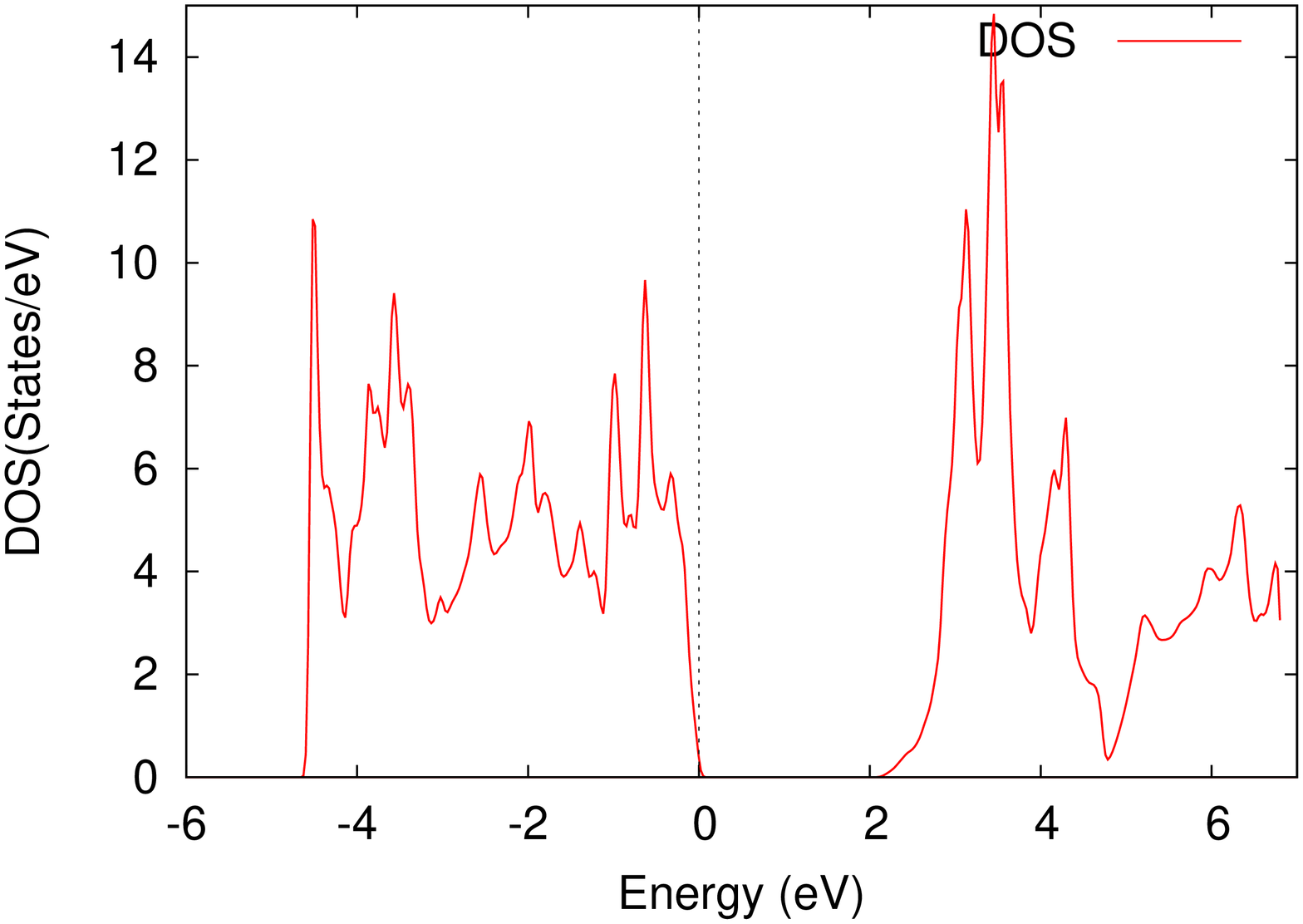}
\caption{(Color Online) Density of states for pristine (a)rutile $TiO_2$ and
(b)anatase $TiO_2$. Total density of states unveils the band gap in both the 
cases.}
\label{fig1}
\end{figure}

\section{Methodology}
 The structures of the three main $TiO_2$ polymorphs are well portrayed 
by the two complementary $Ti_xO_y$ building block. 
Rutile and brookite consists of corner and edge-sharing $TiO_6$ units. 
In rutile, the band gap is about 3.1 eV and in ionic representation each Ti 
atom shares four electron with two oxygen atoms resulting in $Ti^{4+}$ and 
$O^{2-}$. The valance band and conduction band in this wide band-gap 
semiconductor are mainly derived from $O-2p$ and $Ti-3d$ orbitals respectively. 
So when a oxygen vacancy is introduced two electrons are donated to the system. 

The defect induced supercell structures of rutile-$TiO_{2-\delta}$ are relaxed 
on the DFT \cite{dft} level within the generalized-gradient approximation 
using the Perdew-Burke-Ernzerhof (GGA-PBE) functional in the WIEN2K code\cite{wien2k}. The muffin-tin radii, $R_{MT}$ are chosen as 2.0 a.u.
for Ti, and 1.6 a.u. for O. The parameter, $Rk_{max}$ ($Rk_{max}$ stands for 
the product of the smallest atomic sphere radius $R_{MT}$ times the largest k-
vector $k_{max}$ ) is chosen to be 7.0 and 2000 k-points with a $10 \times 10 \times 20$ 
k-mesh is used here for structural optimization. Further, all atoms are 
structurally relaxed until the maximum force is smaller than 0.01 eV/$\AA$. 
The self consistent field (scf) computations are conducted till an energy 
accuracy of 0.001 eV is achieved. The theoretical lattice constants obtained 
from our calculation are in agreement with previous results of parent $TiO_2$.
Then the band structure and the atom-resolved density of
states are calculated from the converged scf calculations.

Here I have used charge self-consistent DFT+DMFT framework which is based on 
the pseudopotential approach for the DFT part and the continuous-time 
quantum-Monte-Carlo method, as implemented in the embedded 
dynamical mean field theory (EDMFTF) package\cite{haule}, for solving the DMFT 
impurity 
problem. I used the GGA+PBE functional form within the Kohn-Sham cycle.
EDMFTF package implements a combined DFT and DMFT derived from the stationary 
Luttinger-Ward functional. In this package the exact double-counting of 
DFT and DMFT, Coulomb interaction and SOC are accomplished nicely and the 
Green’s function is determined self-consistently. DFT+DMFT treatment has been 
successful in explaining theories of strongly correlated electron systems\cite{kim,eug,at1,su,kotliarrmp,supc}. In 
the theoretical model, the non-interacting Hamiltonian is added with the Coulomb 
interaction term, $H_{int}$, \cite{aginter} to incorporate the effects of correlated 
Ti-3d orbitals. A self energy functional($\Sigma_{dc}$) is added to take care 
of the double counting. Thus the total Hamiltonian, except the $\Sigma_{dc}$ term 
is expressed as, 
\begin{equation}
H=\sum_{k,a,\sigma}\epsilon_{k,a}c^{\dagger}_{k,a,\sigma}c_{k,a,\sigma}+U\sum_{i,a}n_{ia\uparrow}n_{ia\downarrow} + $$
$$U'\sum_{i,a,b,\sigma,\sigma'}n_{ia\sigma}n_{ib\sigma'}
\end{equation}

where $\epsilon_{k,a}$ is the band dispersion and $U$ and $U'$ are
the intra- and inter-orbital Coulomb interaction terms between electrons with 
opposite spins of a orbital and between electrons with same spins in two orbitals respectively. The coulomb interaction terms are varied over 
a realistic range to get nice agreement with earlier experi-
ments on TiO$_2$ and the total Hamiltonian is solved using
the DMFT method.
The correlated orbitals are then handled dynamically within the DMFT based on orbital 
projection-embedding scheme achieved via the EDMFT package. The impurity solver used 
in this DMFT code is the continuous time quantum Monte Carlo(CT-QMC)
in the hybridization expansion method\cite{haule1}. The parameters, namely, the Coulomb 
interaction $U$, and the inverse temperature, $\beta(= 1/k_BT)$ are varied
within an experimentally realizable range to get the behavior 
of the oxygen vacancy in the system. The DFT+DMFT
calculations are converged with respect to the charge density, 
chemical potential and the self energy. Finally the maximum entropy method \cite{jarrell} is 
used for the analytical continuation of the self-energy from the imaginary 
axis to real frequencies with an auxiliary Green’s function. Then
from the real frequency Green’s function, the momentum resolved spectral functions and the density of states are obtained.

\begin{figure}
(a)
\includegraphics[angle=270,width=0.9\columnwidth]{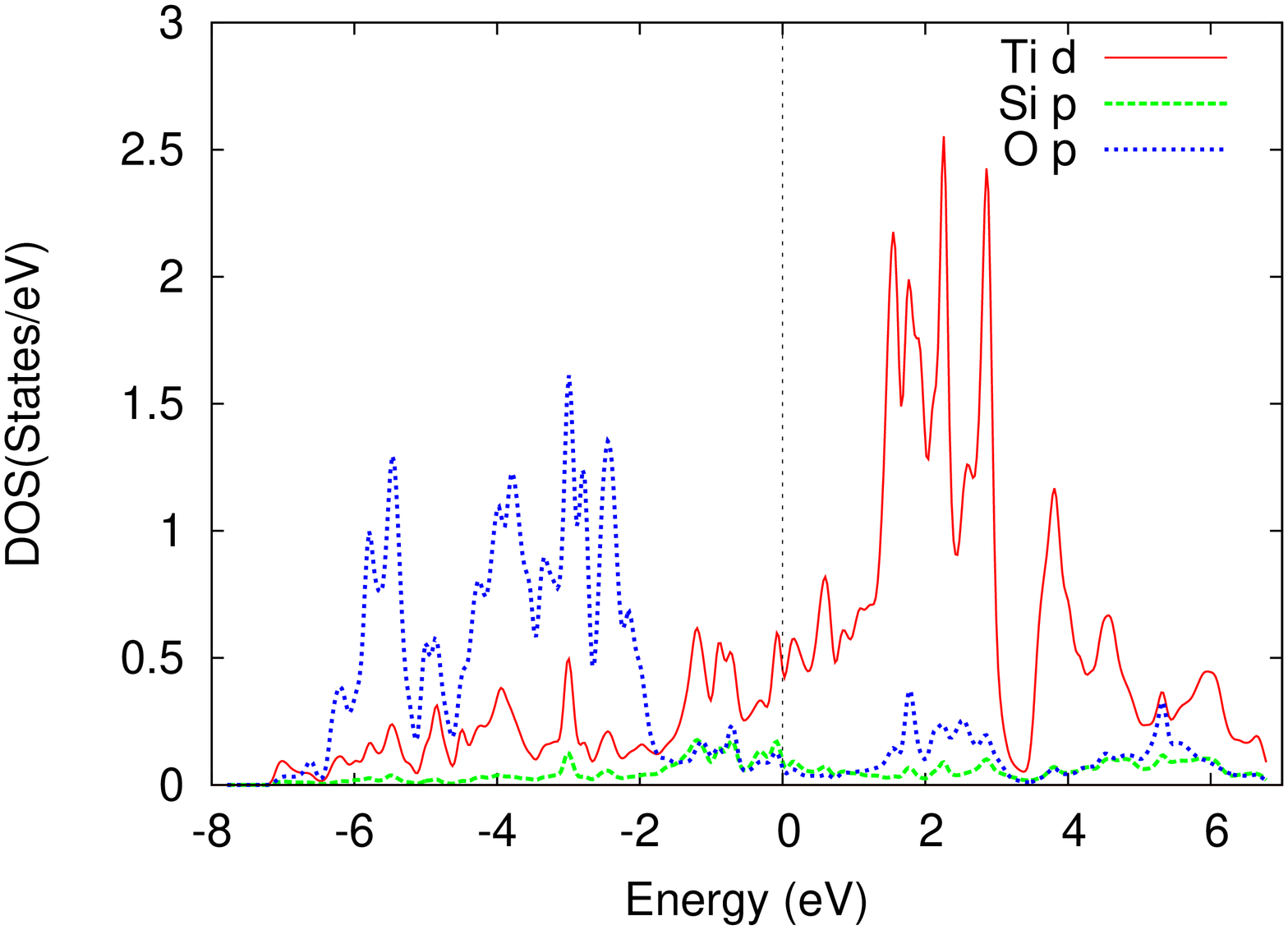}

(b)
\includegraphics[angle=270,width=0.9\columnwidth]{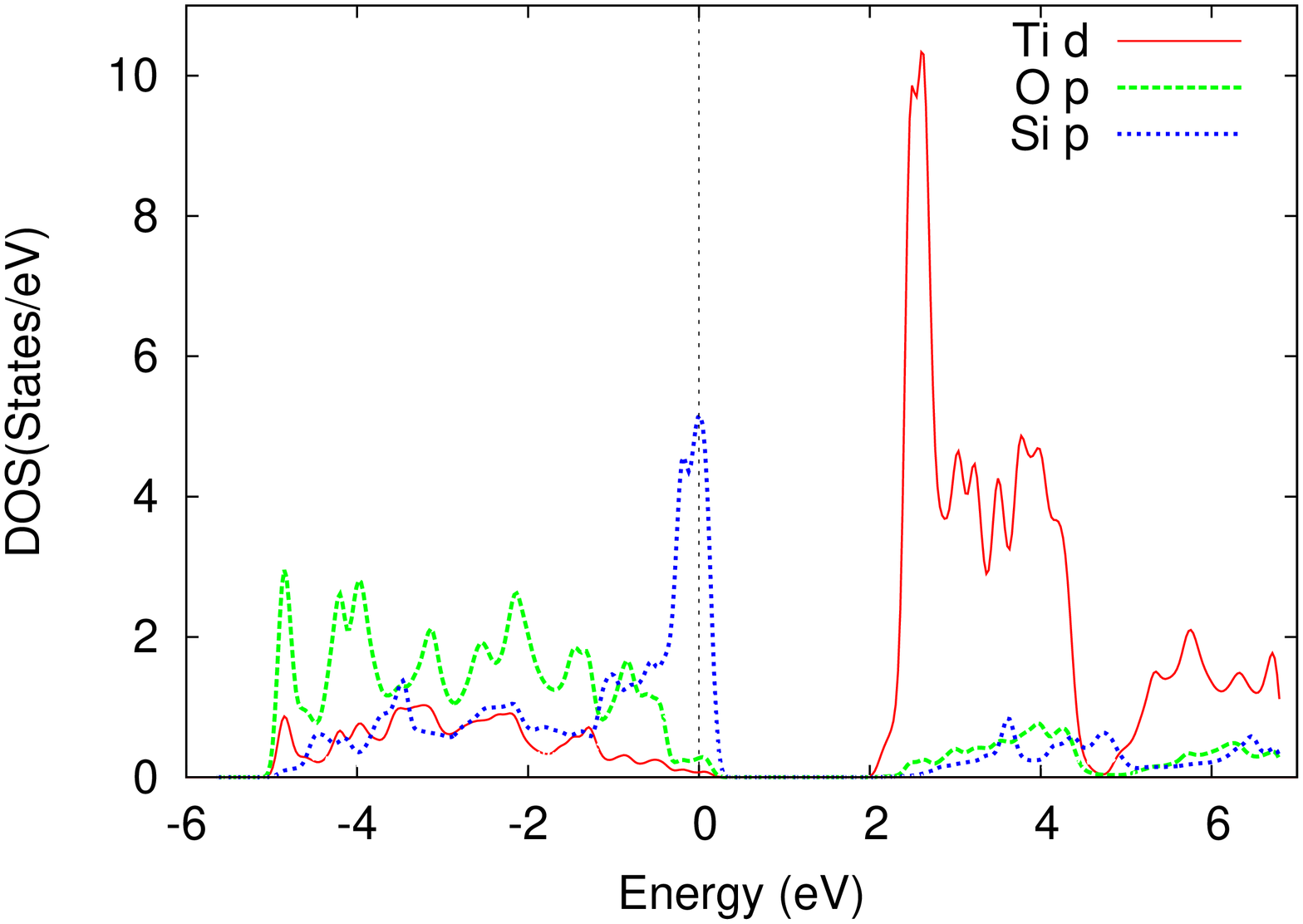}
\caption{(Color Online) Density of states for doped (a)rutile $TiO_2$ and
(b)anatase $TiO_2$. Doping is induced by Si to engineer oxygen vacancy in $TiO_2$. Density of states plots reflects the band gap is closed in both the 
cases and Fermi level has finite weight of impurity levels and Ti-d band.}
\label{fig2}
\end{figure}
\begin{figure}
(a)
\includegraphics[angle=270,width=0.9\columnwidth]{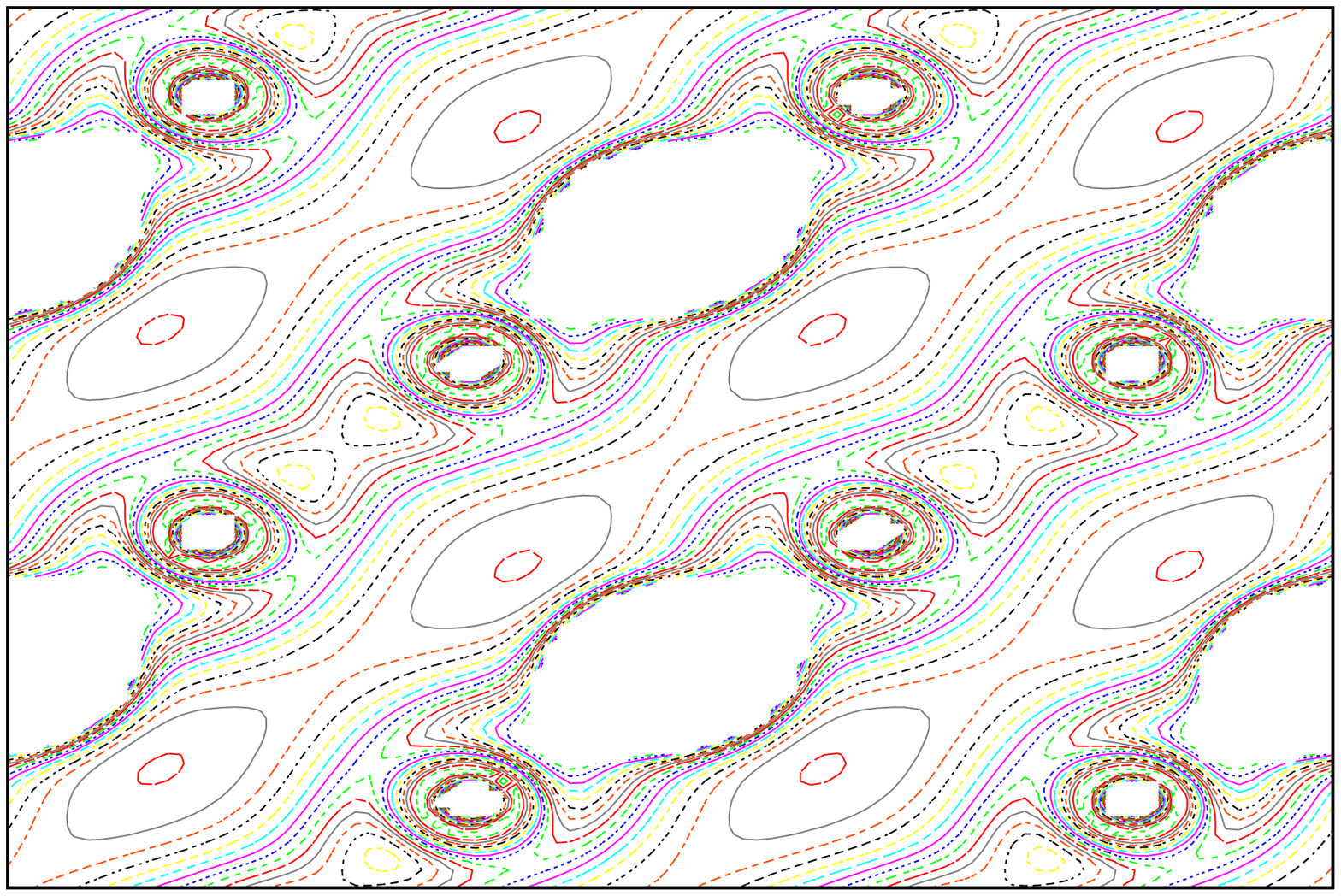}

(b)
\includegraphics[angle=270,width=0.9\columnwidth]{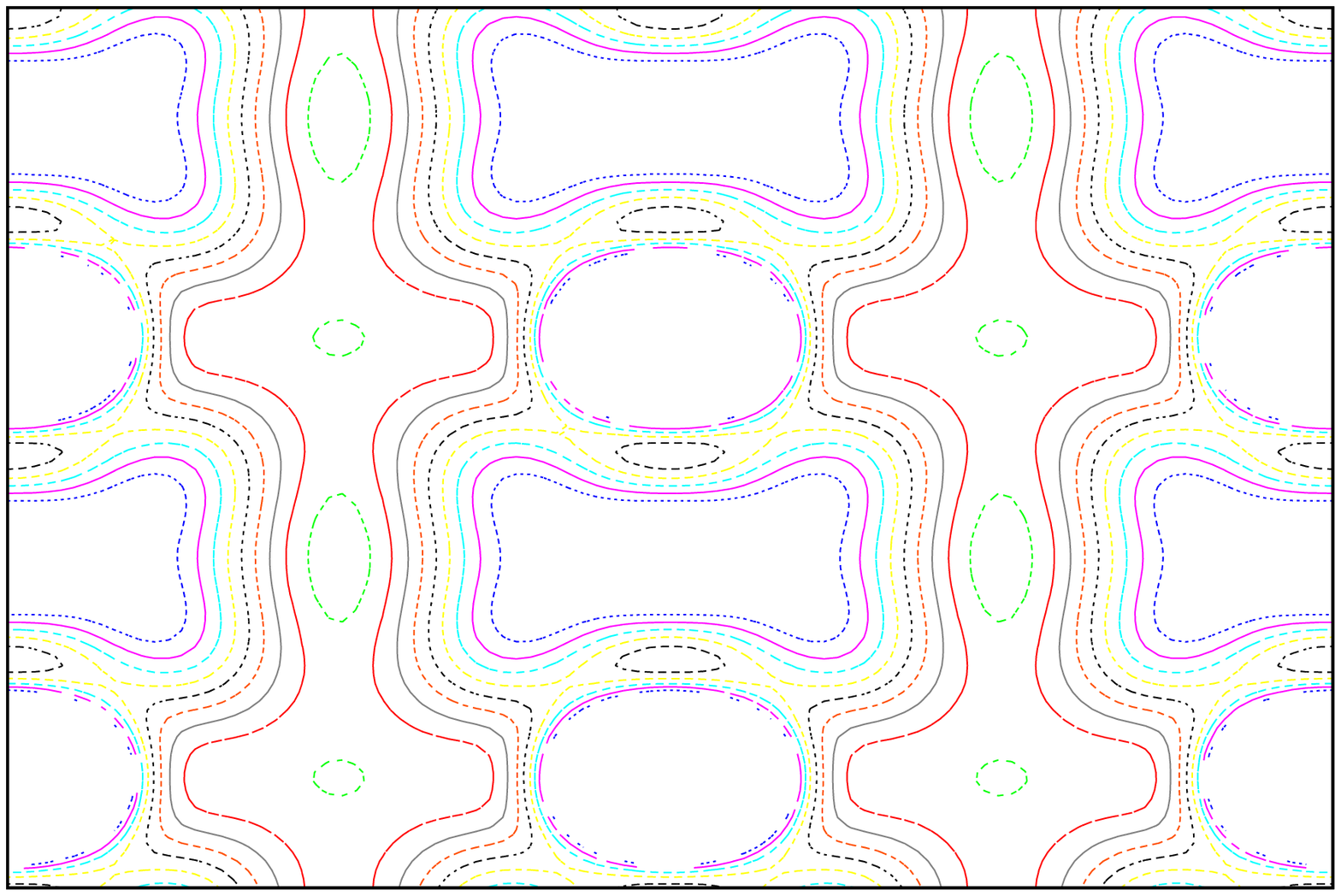}
\caption{(Color Online) (a) Electronic charge density plot for (a) rutile and (b) anatase phase of $TiO_2$ in the plane containing $Ti, O, Si$. Charge density 
plots shows more conducting region in comparison to parent $TiO_2$.}
\label{fig3}
\end{figure}
\begin{figure}
(a)
\includegraphics[angle=270,width=0.9\columnwidth]{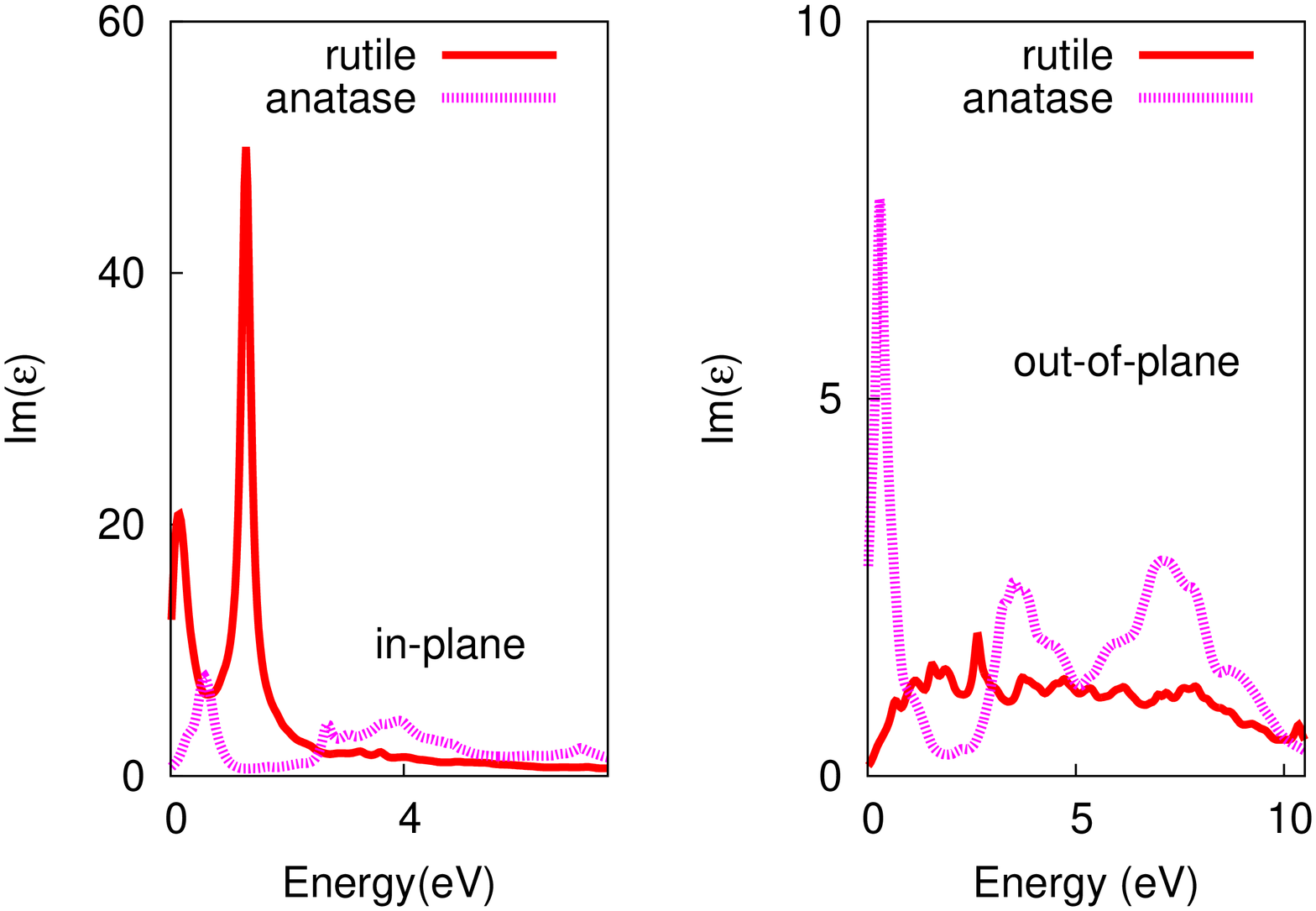}

(b)
\includegraphics[angle=270,width=0.9\columnwidth]{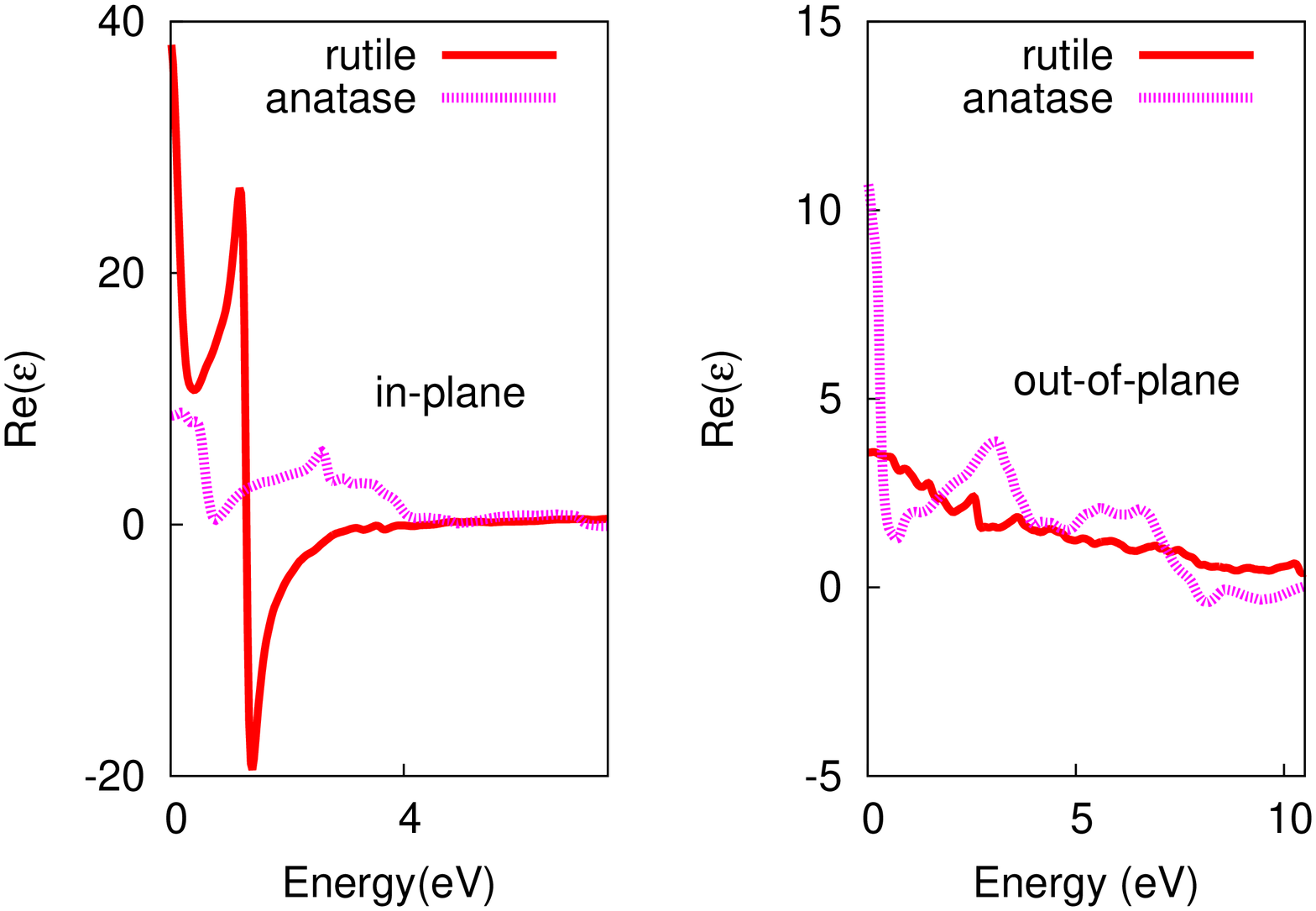}
\caption{(Color Online) The (a) imaginary and (b) real part of the dielectric 
function as calculated from DMFT for rutile and $TiO_2$ along in-plane and 
out-of-plane direction The static value of real part of dielectric function has 
increased due to oxygen vacancy and absorption peak is also starting from zero 
energy making it more efficient for solar cell application.}
\label{fig4}
\end{figure}
\begin{figure*}
(a)
\includegraphics[angle=270,width=0.58\columnwidth]{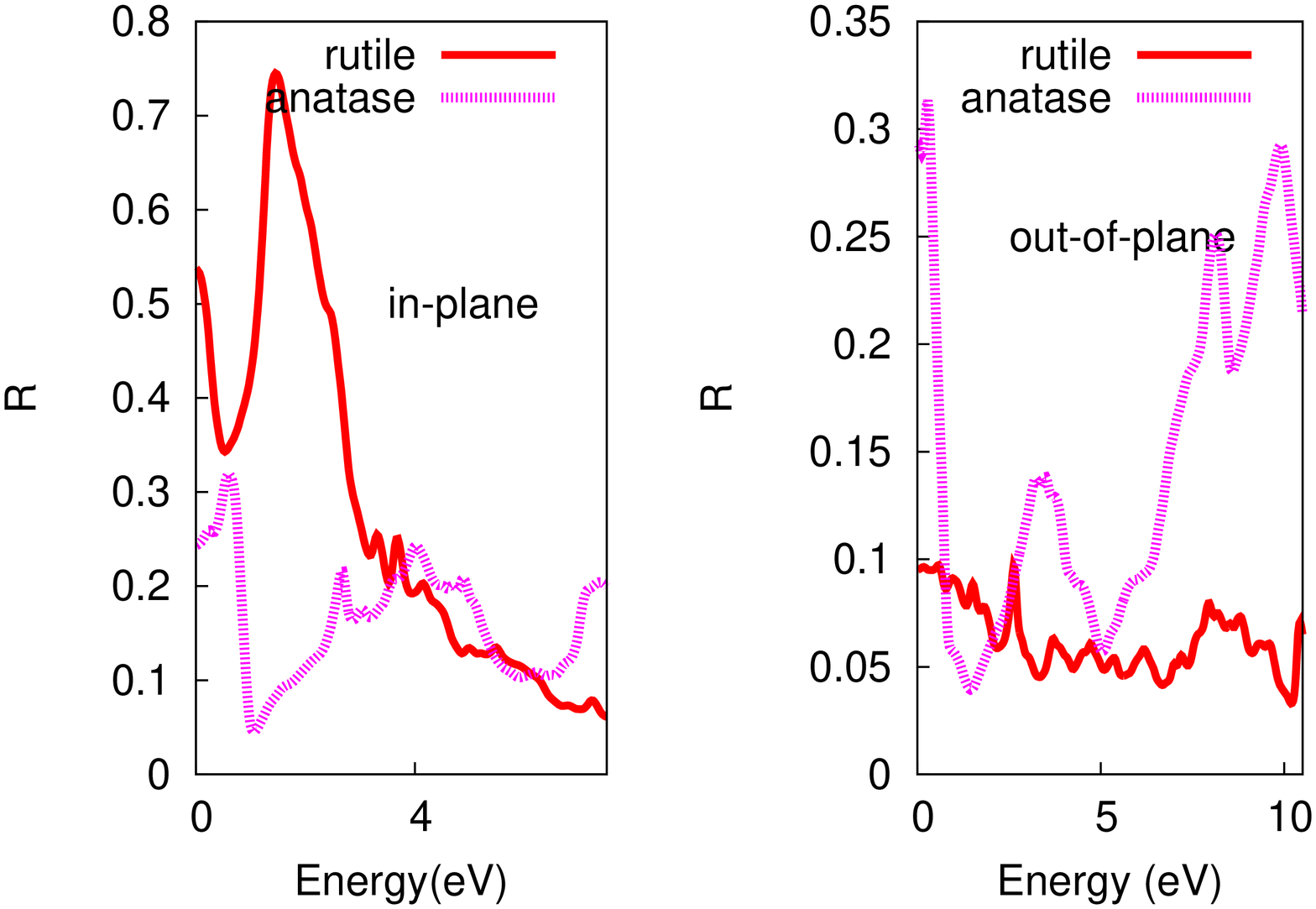}
(b)
\includegraphics[angle=270,width=0.58\columnwidth]{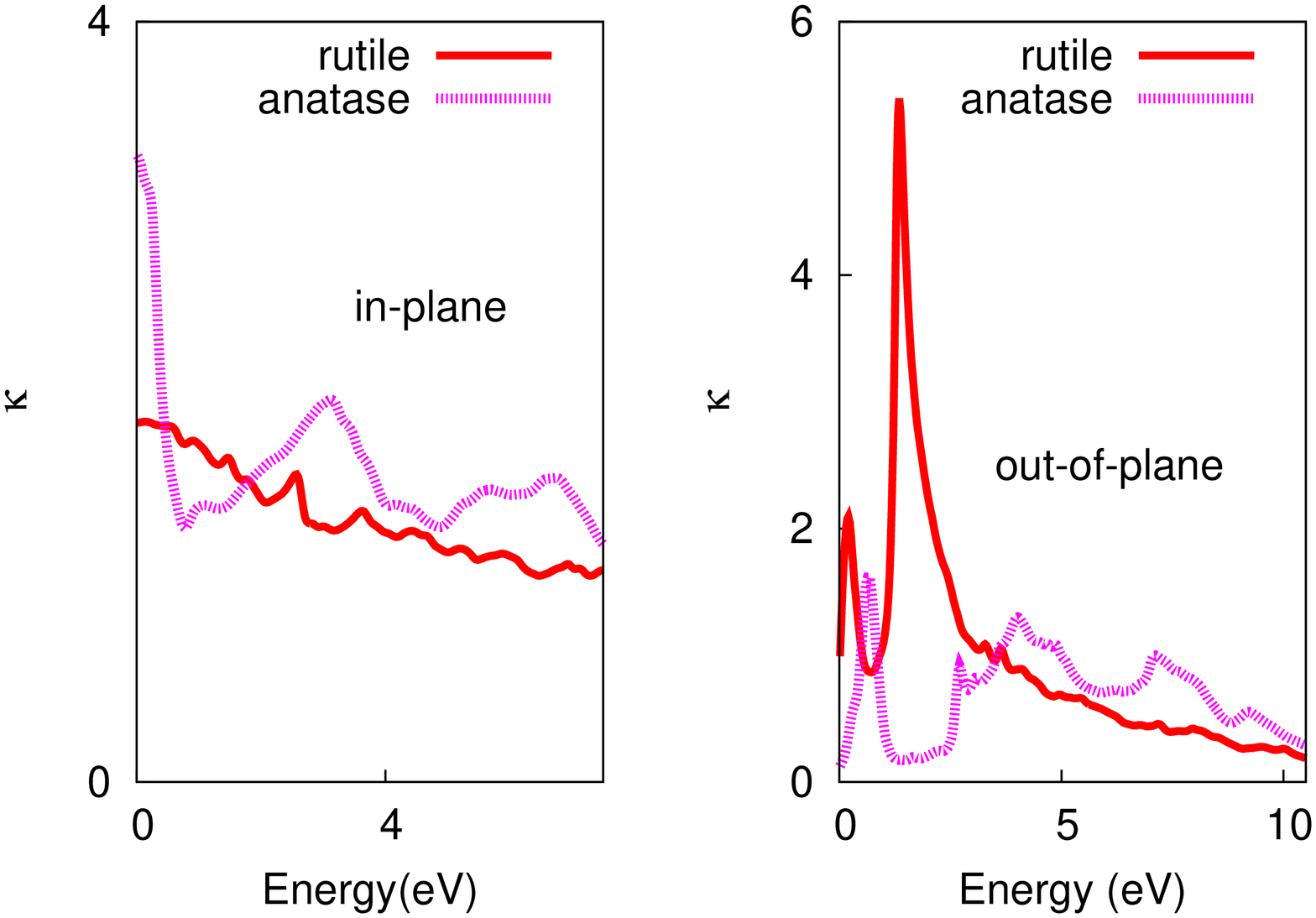}
(c)
\includegraphics[angle=270,width=0.58\columnwidth]{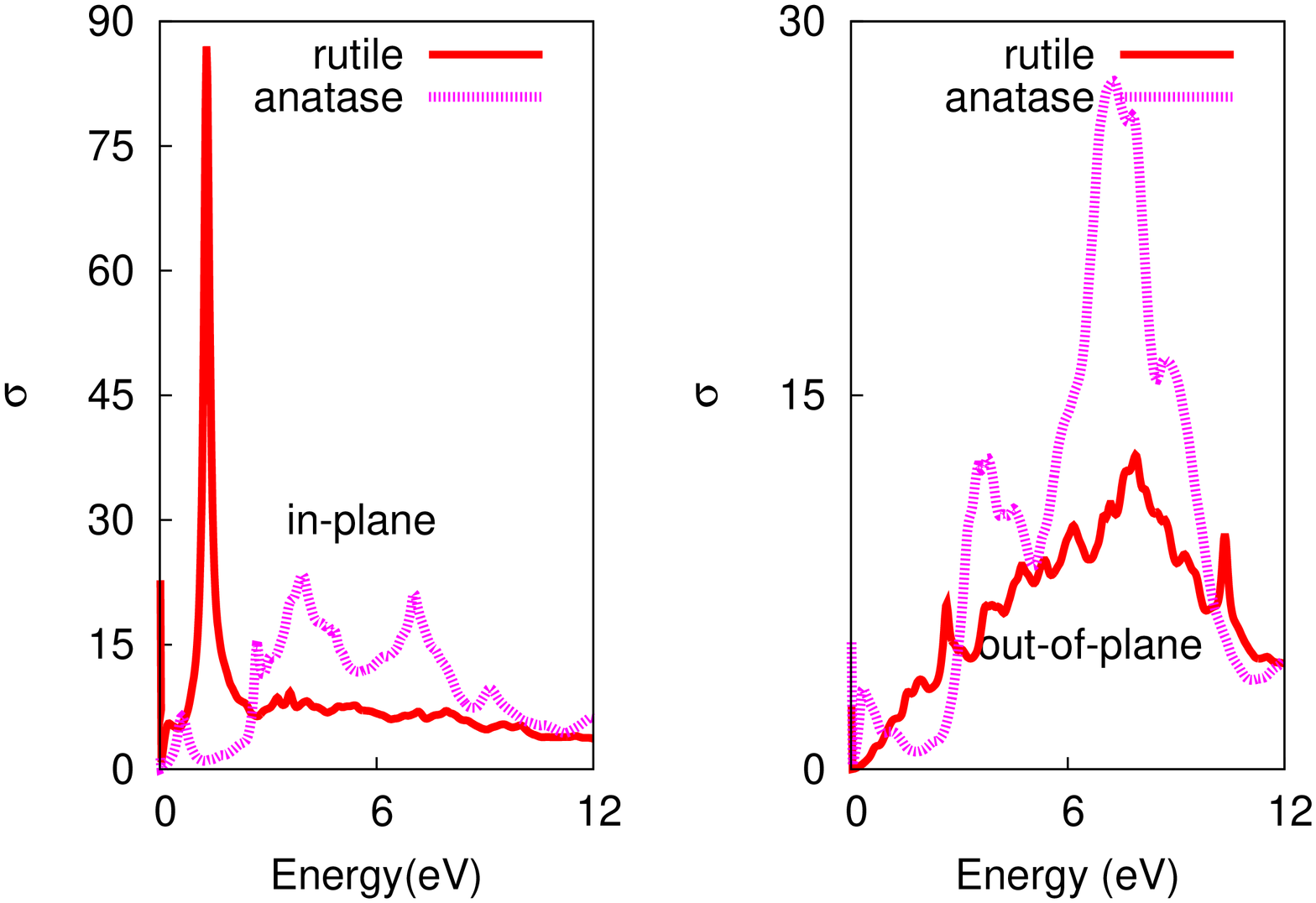}

(d)
\includegraphics[angle=270,width=0.58\columnwidth]{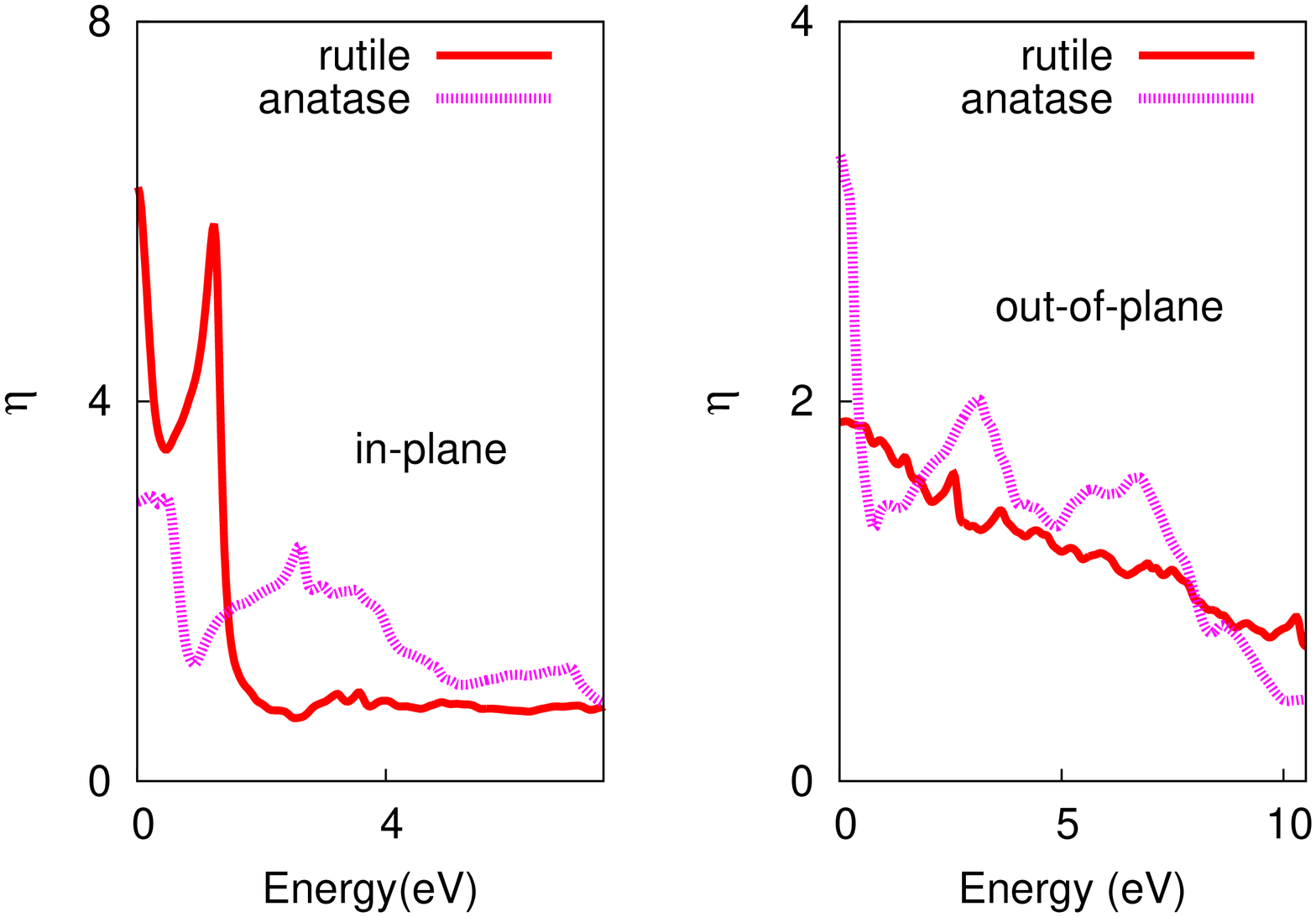}
(e)
\includegraphics[angle=270,width=0.58\columnwidth]{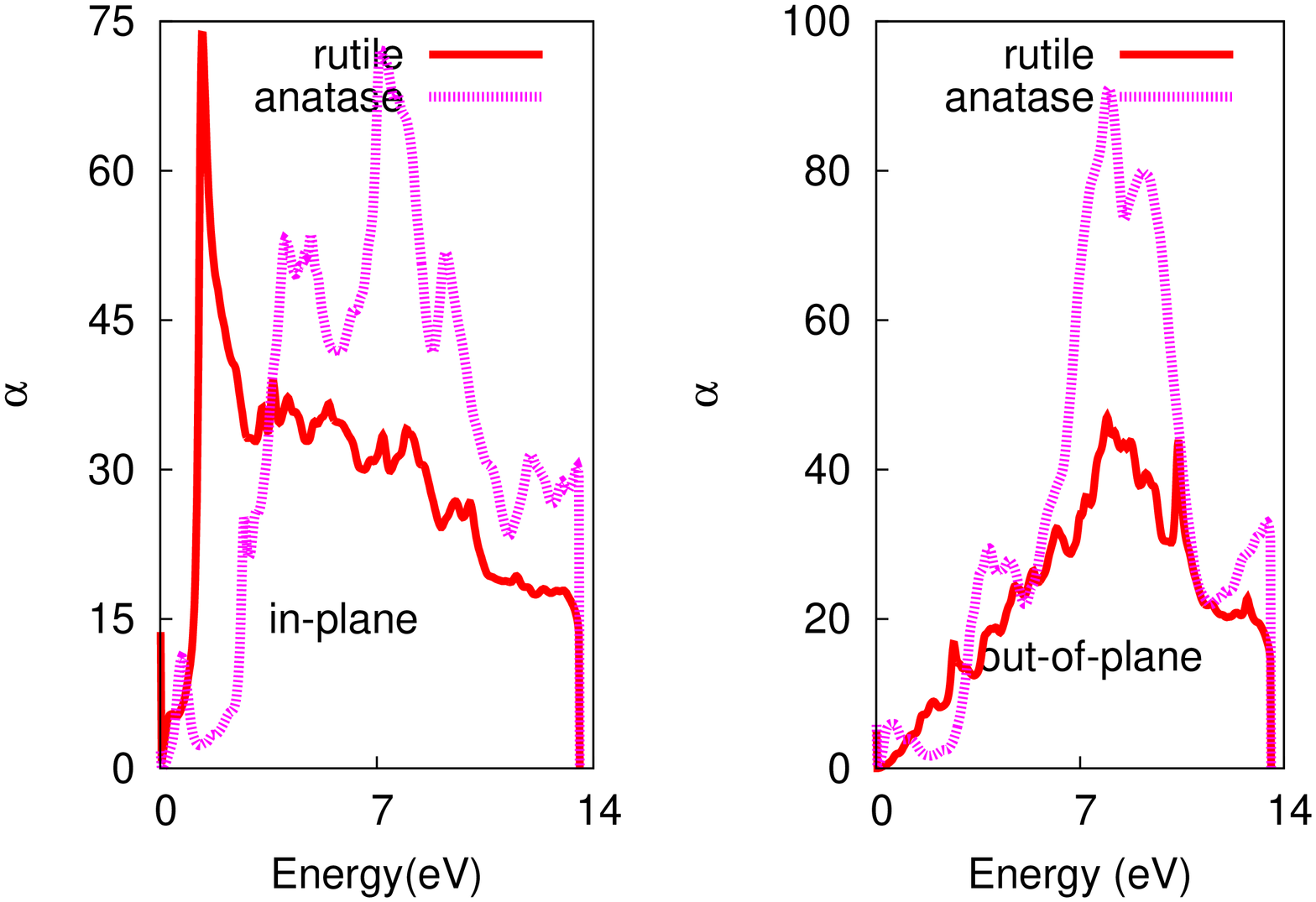}
(f)
\includegraphics[angle=270,width=0.58\columnwidth]{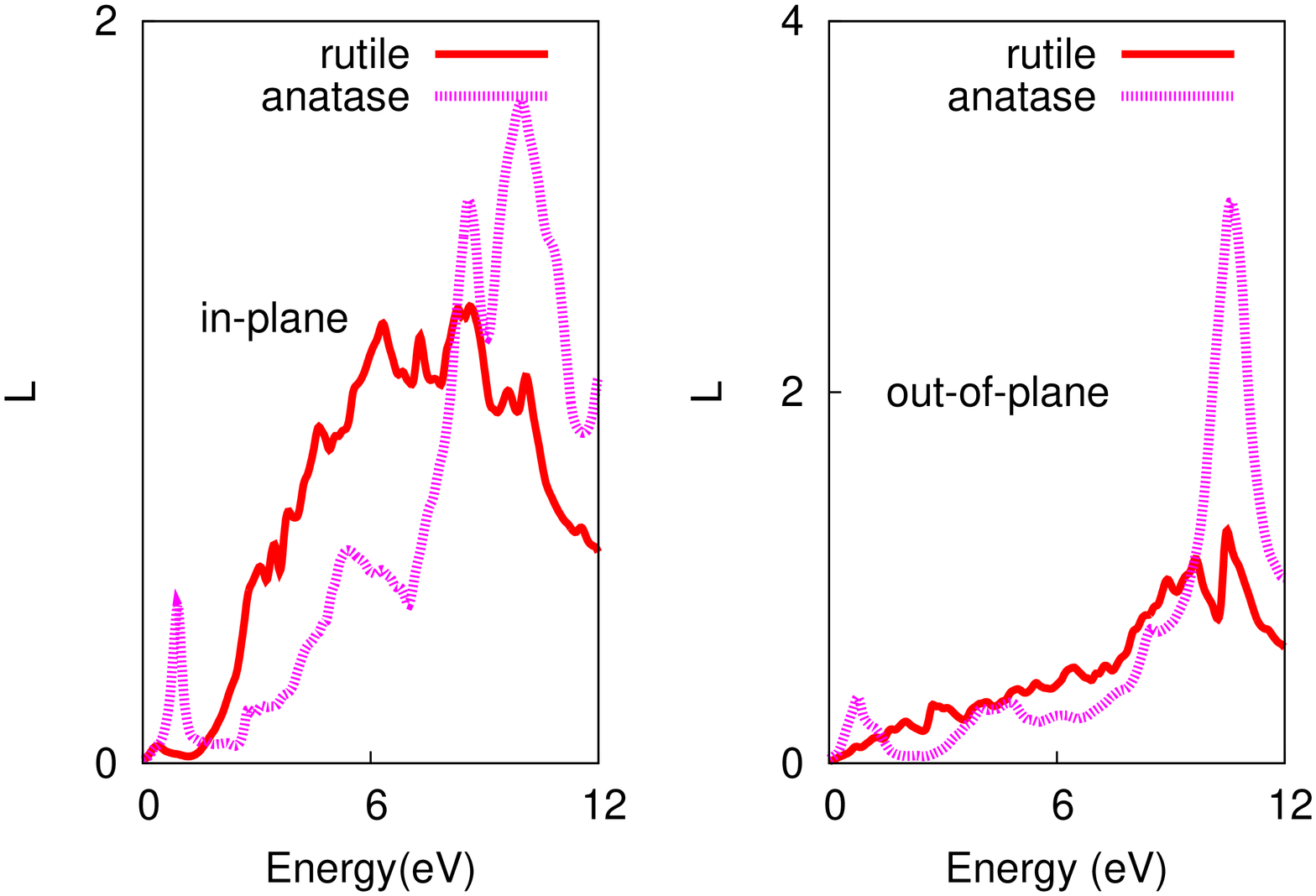}
\caption{(Color Online) The DMFT optical constants plot: (a)reflectivity, 
(b)extinction coefficient, (c) optical conductivity, (d)refractive index, 
(e)absorption coefficient and (f) dielctric loss spectra for rutile and anatase
$TiO_2$ crystal structure. Optical constants are plotted for in-plane and 
out-of-plane direction.}
\label{fig5}
\end{figure*}

\section{Results}
The choice of coulomb interactions ($U$ and $U'$) has an 
important role for getting the electronic properties of the doped transition metal oxide with 
vacancy. For rutile and anatase $TiO_2$ I have chosen to calculate for a 
reasonable range of $U$ values and found that for $U=8.0$ eV and $U'=6.8$ eV 
the band gap energies exactly agrees well with the experimental results\cite{3,5,6}. The 
orbital dependent density of states (DOS) for pristine rutile and anatase $TiO_2$ 
is shown in Fig.1. Though theoretical and experimental band gap have slight 
difference but that should not affect the electronic structure and transport 
properties of defect induced $TiO_2$ polymorphs. 
To create the outcome of oxygen deficiency a realistic model with one oxygen atom 
eliminated from the $4\times4\times6$ supercell of anatase and rutile $TiO_2$ is 
formed. In Fig.2 the orbital dependent DOS for the doped rutile and 
anatase $TiO_2$ with oxygen vacancy are displayed. Each $TiO_2$ is having 
oxygen atoms and adjacent 
Ti atoms. In rutile structure among Ti atoms two are symmetrically 
equal so Ti-O bond length is somewhat different. Removal of an oxygen atom
 effects in three Ti bonds (with d character) pointing in the direction of
 the oxygen vacancy position. However the vacancy position is occupied by two 
electrons.
The resultant orbital density of states show impurity states predominantly
 emerge from the
$Ti-d$ orbitals and doped atoms ($Si$ is used here) and new states develop in the band gap. 
The symmetrically equivalent $Ti$ atoms will have same DOS (not shown here). 
The partial charge density  distribution is also shown in Fig.3 corresponding 
to the defect states in both rutile and anatase $TiO_2$.
The partial charge density distribution shows that vacancy of  
one oxygen atom in both the $TiO_2$ lattice develops a highly conducting 
region in real space, and the Ti atoms that were primarily having bond
with the oxygen become localized in the vacanct position. Due to the 
impurity doping 
new energy levels come into the band gap which indicates that compared with 
both the pristine $TiO_2$, there is extra charge on the O-deficient Ti ions 
in the presence of an oxygen vacancy. As an outcome of doping the O-deficient 
Ti ions transform from higher valence state to a lower valence state after 
creating a neutral oxygen vacancy.
Now since anatase $TiO_2$ has wide applications in the field of optoelectronic 
devices and solar cell conversion efficiency so next I will show optical 
properties of oxygen deficient $TiO_2$. The imaginary part of the dielectric 
function is defined as optical absorption. $TiO_2$ has tetragonal structure 
and literature shows it exhibits optical anisotropy. Hence, the presence of 
anisotropy is also obvious for the doped systems, and it is more pronounced in 
the low energy part of the spectra. So I present optical properties in both the 
in-plane and out-of plane direction. 

The absorption spectra (Fig.S1(Supplementary)) of undoped rutile structure 
shows 3.0 eV as initial 
energy for absorption from DMFT calculations while dielectric function at zero energy is having constant value of 5.75 and 5.55 along in-plane and out-of-plane 
direction which is consistent with earlier results on rutile $TiO_2$\cite{45,56}. 
So our calculation with DFT+DMFT spectra shows very good agreement with earlier 
results and the precise absorption spectra from oxygen vacant states also can 
be achieved.
Similarly for anatase structure the onset of absorption starts from 2.4 eV and 
real part of dielectric function is having values 5.35 and 5.2. 
Now I will show optical absorption spectra which carry a number of important 
properties like refractive index($\eta$), relectivity (R), optical conductivity
($\sigma$), extinction coefficient($\kappa$) and optical loss spectra. In 
DFT+DMFT from EDMFTF package all of these can be calculated. The theoretically
calculated in-plane and out-of-plane refractive indices for rutile $TiO_2$ are 
in close agreement with the experimental results\cite{56}. Therefore this 
DFT+DMFT method is good enough to produce the doped $TiO_2$ case.

The many body theory in rutile $TiO_2$ is recently reported\cite{58} but for 
anatase structure it is still not reported. DMFT can unveil the presence of 
defects which affects the optical properties of pristine $TiO_2$ structures.    
Fig.4a and Fig.4b shows real and imaginary part of dielectric function of 
doped rutile and anatase $TiO_2$. In both the case though the nature of the
absorption pattern is same but in pristine $TiO_2$ there was no absorption 
peak below 3 eV but in doped cases it is different and absorption peaks start 
from almost zero energy. However like parent structure optical anisotropy 
is present in the doped system also and it is pronounced in both the structure. 
Though presence of low energy peaks in both the case reflect the two directions 
as optically active for polarization of light. The reason for optical activity 
can be explained from the band structure itself. Since oxygen vacancy is created
 and the system is doped with Si, which is having a tendency to accept negative charges so the 
Fermi level will be changed towards conduction band minima. So here Si will act as 
n-type dopant and $p_x, p_y$ and $p_z$ orbitals actively participate in optical 
transitions at low energy. Thus n-type dopants make rutile and anatase $TiO_2$ 
optically active in low energy for light polarized along $xy$ and $z$ direction.

Now Fig.5 shows optical constants in in-plane and out-of-plane for the enrgy 
range 10 eV which is reasonable since I have taken DFT bands also in this range. 
The refractive index is having a peak (Fig.4d) in the infrared and visible range followed
 by a fast decrease in the ultraviolet range for rutile structure when 
electric field is in-plane while there is a slow decrease for out-of plane 
direction. For anatase structure refractive index follow opposite behaviour, with
 two small peaks in in-plane visible and infrared range and a slow decrease 
in the UV range while out-of plane refractive index is having a large peak in 
infrared region and decreasing fast in the UV region. Depending on the 
structure refractive index is near about 2 in the visible range of the 
electromagnetic spectrum.  A similar behaviour is found for extinction 
coefficient (Fig.4b), which is responsible for attenuation of electric field 
and is a cause of dielectric loss. Next the dielectric loss is presented in Fig.4f 
which shows highest loss in low wavelength region. According to all these 
results optical constants can be changed by oxygen vacancy or doping. Next I 
have theoretical optical conductivity also from DMFT. Optical conductivity 
in turn is related to refractive index (Fig.4d), absorption coefficient (Fig.4e)
 and light speed. Optical conductivity is the electrical conductivity as a 
result of movement of the charge carriers due to alternating electric field 
generated by electromagnetic wave. It is observed that optical conductivity 
increases in doped $TiO_2$ due to the new energy levels in the bad gap. The
impurity levels in the band gap facilitate the electrons from the valance 
band transmits to the conduction band.

In conclusion, the electronic structure of rutile and anatase $TiO_2$ has been 
investigated using DFT+DMFT method with the introduction of oxygen vacancies and Si doping. 
Electronic charge density shows formation of a high conducting region in 
both $TiO_2$. The theoretical results are consistent with earlier findings of 
n-type dopant in $TiO_2$\cite{arxiv10569}. Optical excitations are found to be 
modified by oxygen vacancy while they remain anisotropic as in the parent system. Optical activity 
in both in-plane and out-of-plane direction is found for doped rutile and 
anatase structure. As a result of defect and doping optically allowed transitions
are from very low energy which improves its applicability in opto electronic 
and solar absorption properies.

\section{ACKNOWLEDGEMENT}
S. Koley acknowledges DST women scientist grant
SR/WOS-A/PM-80/2016(G) for finance and also thank
Prof. M C Mahato for mentoring and useful conversa-
tions.

\pagebreak
\pagenumbering{arabic}
\setcounter{figure}{0}
\section{Supplementary Information}
\renewcommand{\thepage}{S\arabic{page}}
\renewcommand{\thesection}{S\arabic{section}}
\renewcommand{\thetable}{S\arabic{table}}
\renewcommand{\thefigure}{S\arabic{figure}}
\begin{figure}
(a)
\includegraphics[angle=0,width=0.3\columnwidth]{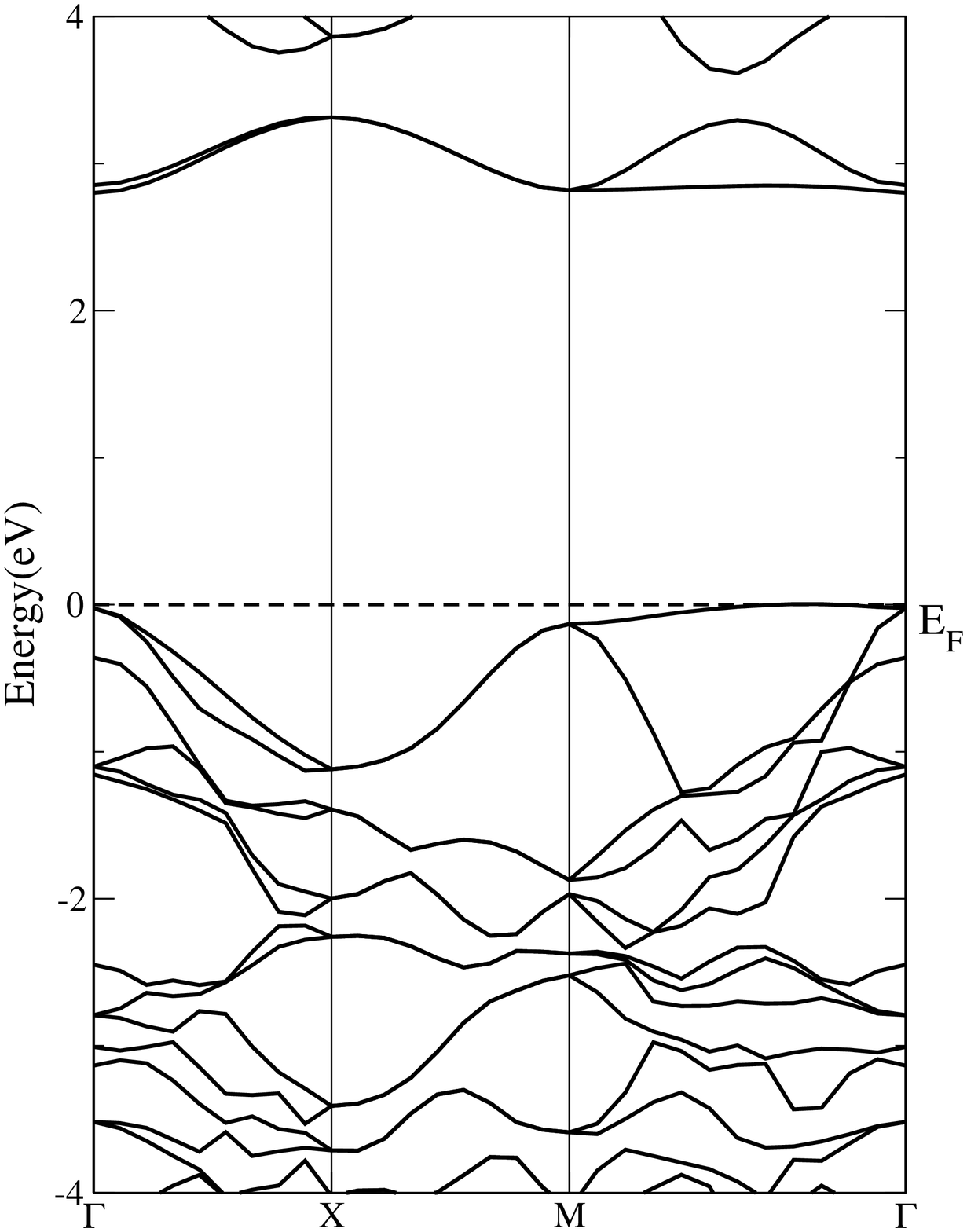}
(b)
\includegraphics[angle=0,width=0.3\columnwidth]{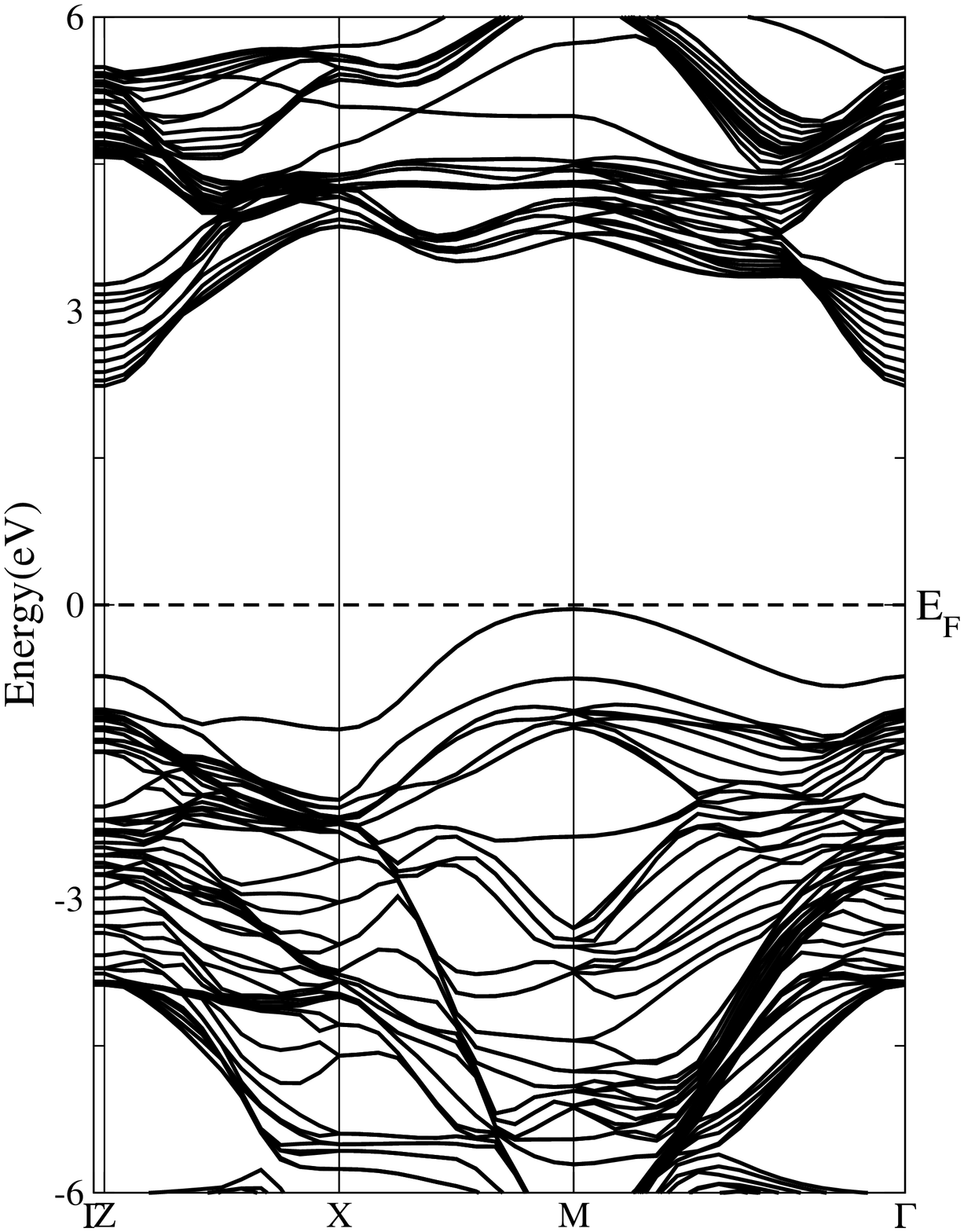}

(c)
\includegraphics[angle=0,width=0.3\columnwidth]{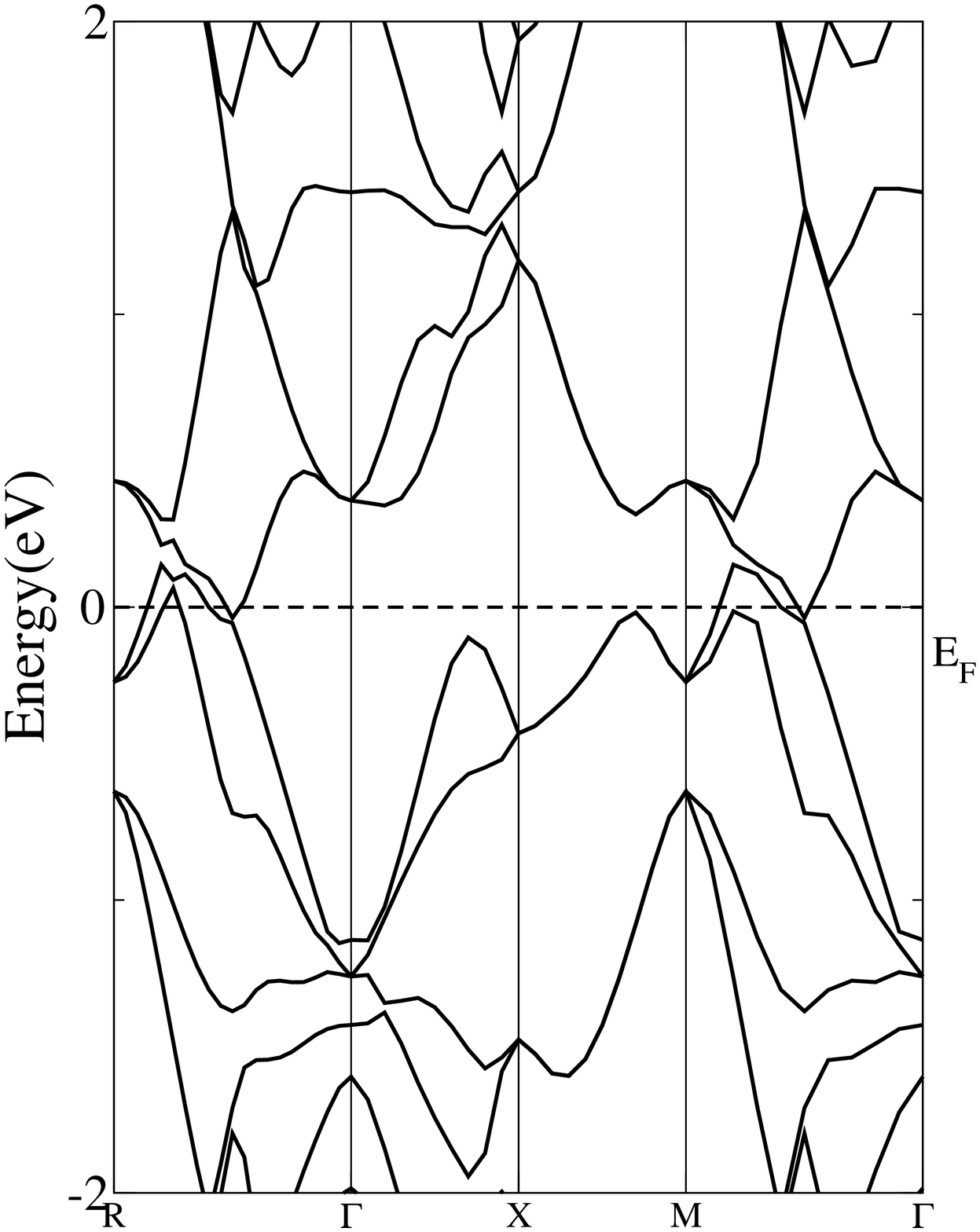}
(d)
\includegraphics[angle=0,width=0.3\columnwidth]{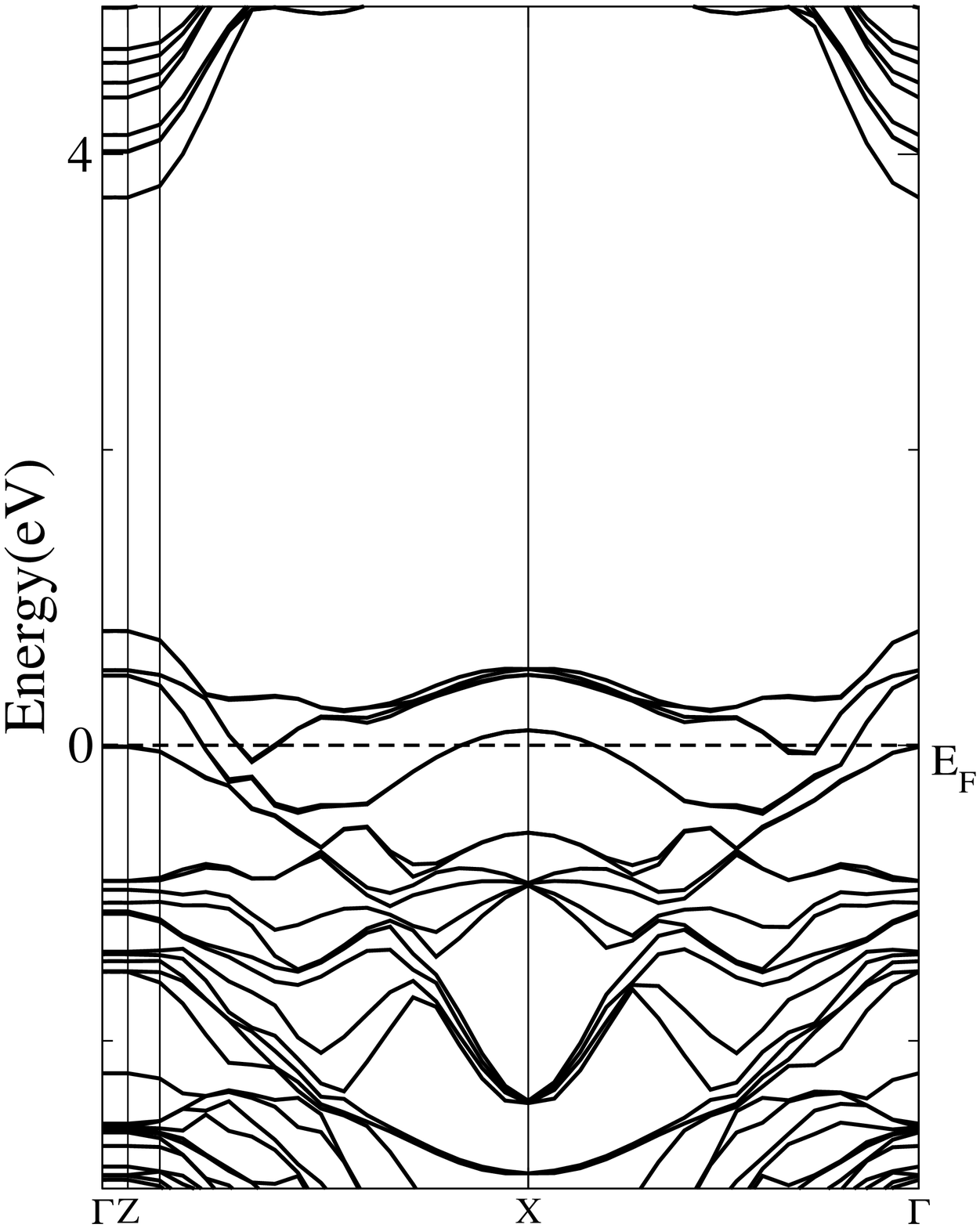}
\caption{(Color Online) The plots show band structure of pristine (a)rutile and 
(b)anatase and oxygen deficient (c) rutile and (d) anatase $TiO_2$ structure.}
\label{figs1}
\end{figure}
\begin{figure}
(a)
\includegraphics[angle=270,width=0.5\columnwidth]{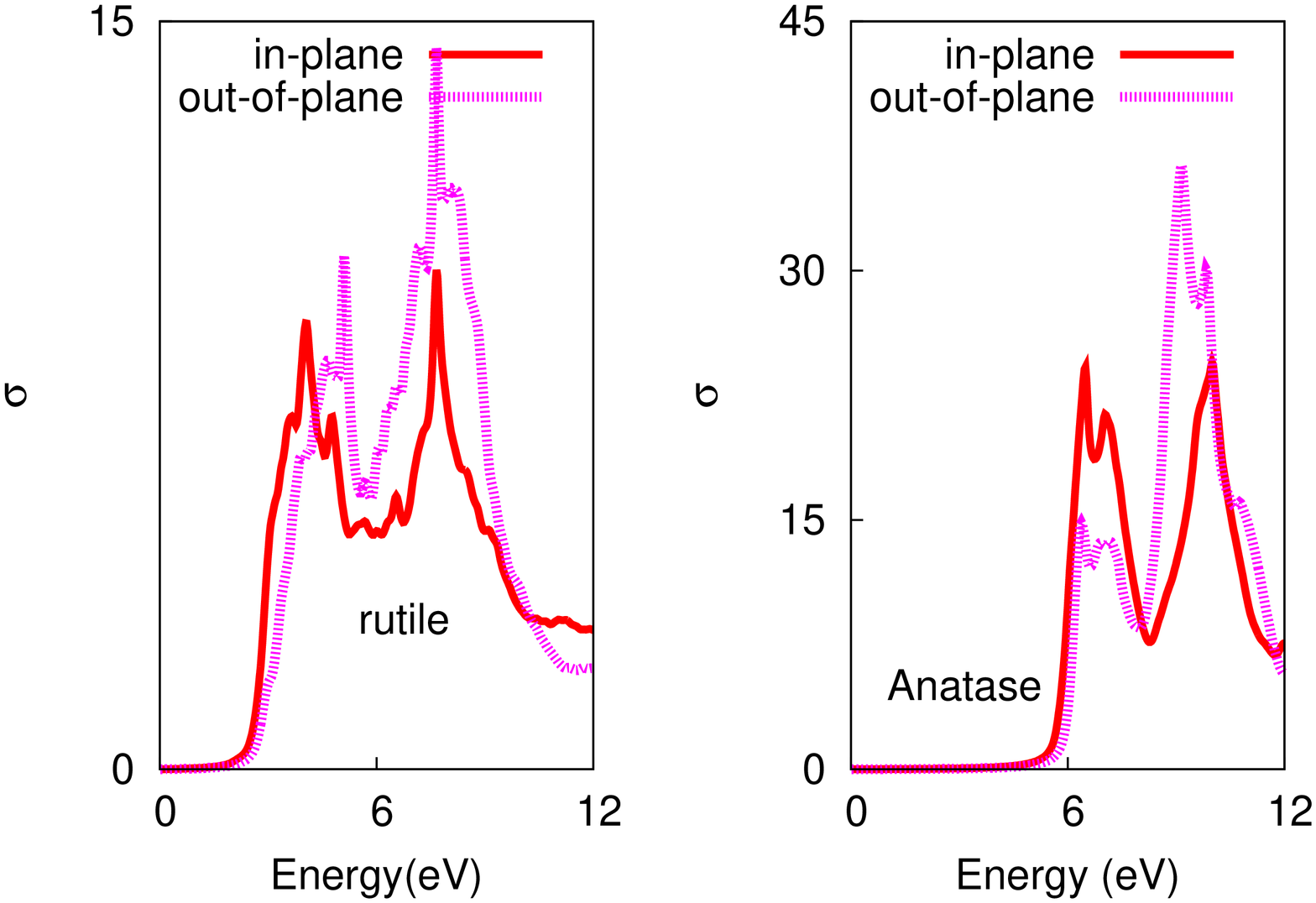}

(b)
\includegraphics[angle=270,width=0.5\columnwidth]{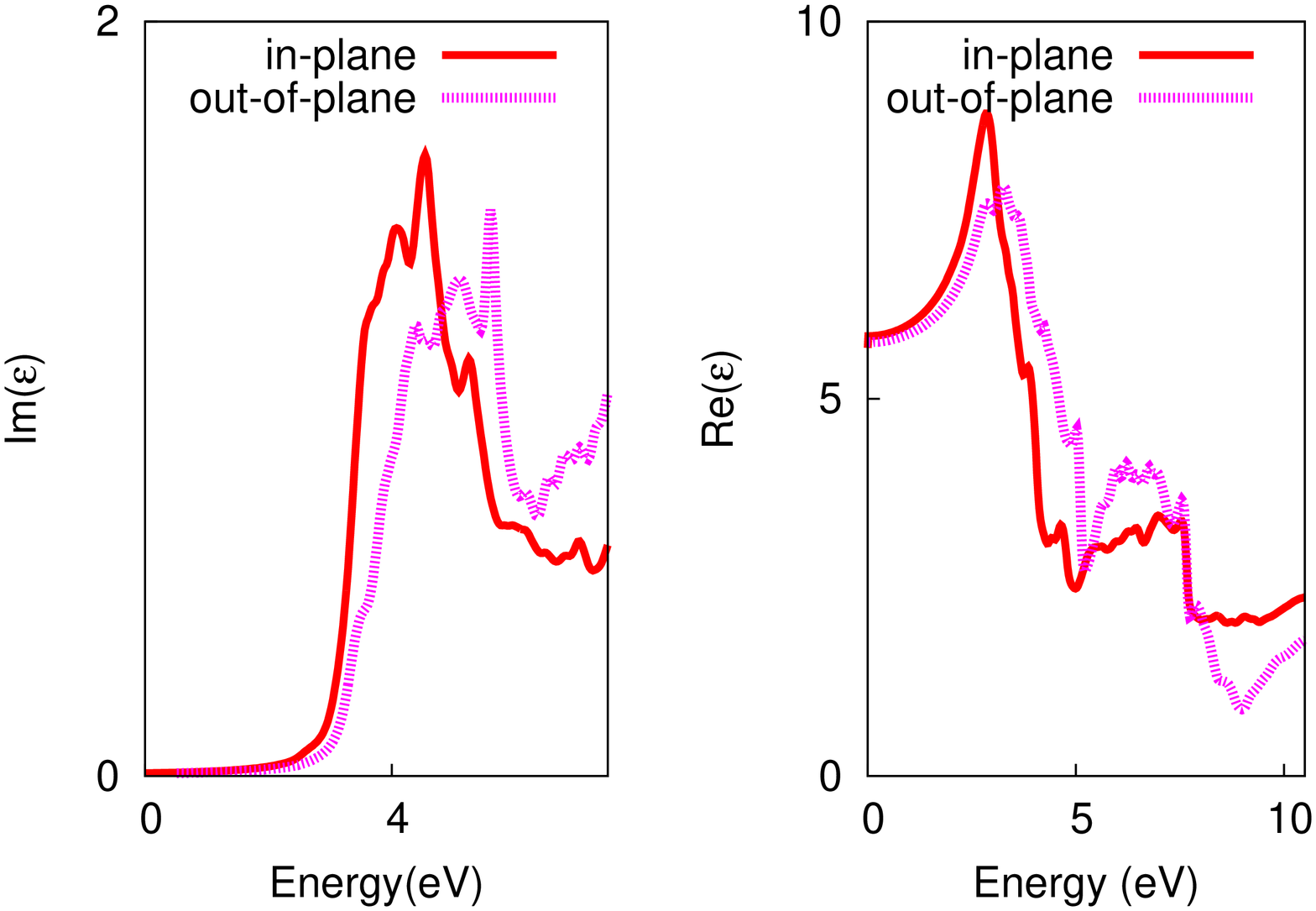}

(c)
\includegraphics[angle=270,width=0.5\columnwidth]{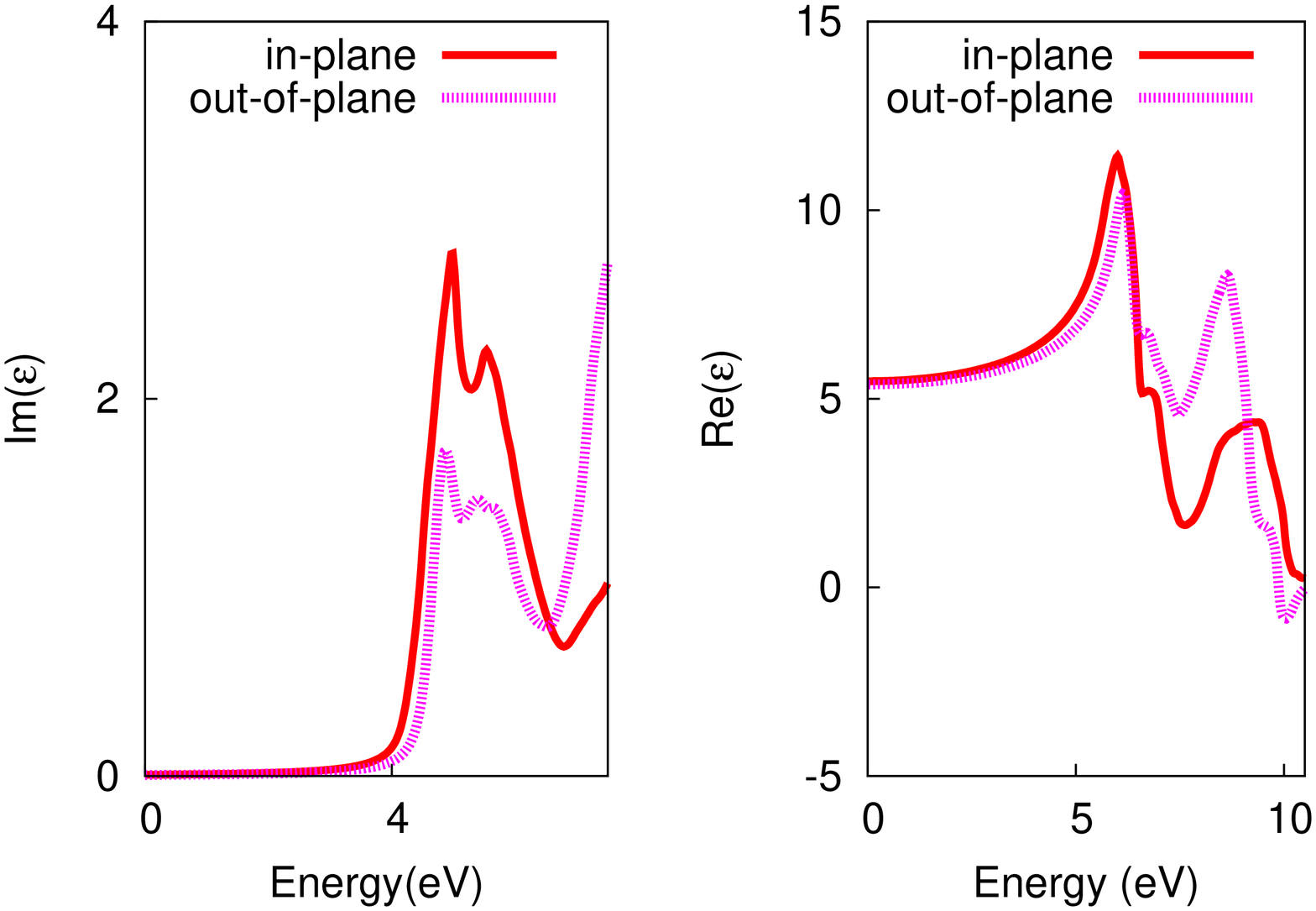}
\caption{(Color Online) (a) Optical Conductivity, (b) dielectric function of rutile and (c) dielectric function of anatase of parent $TiO_2$ structure.}
\label{figs2}
\end{figure}
\begin{figure}
(a)
\includegraphics[angle=270,width=0.5\columnwidth]{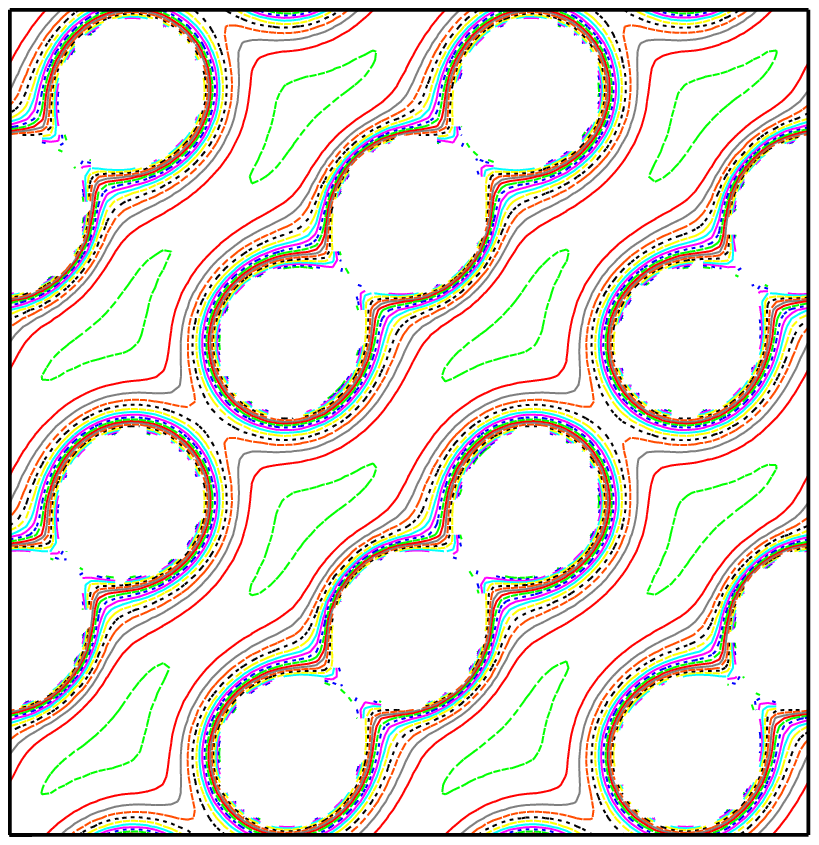}

(b)
\includegraphics[angle=270,width=0.5\columnwidth]{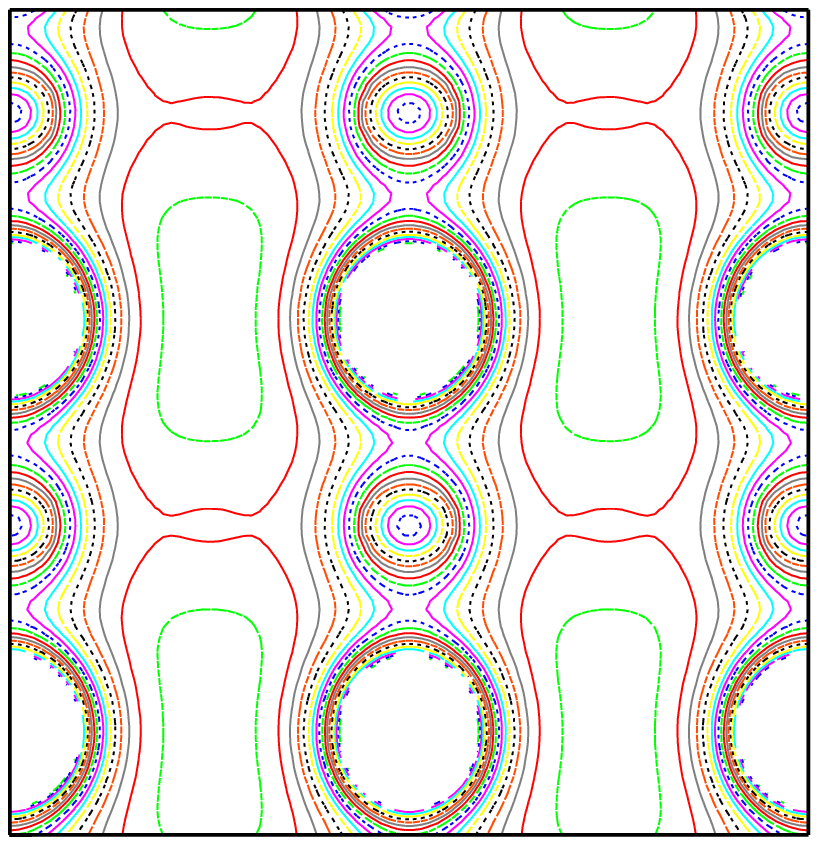}
\caption{(Color Online) Electronic charge density of pristine (a) rutile and (b) 
anatase $TiO_2$ structure in the plane containg $Ti-O$.}
\label{figs3}
\end{figure}  

\begin{thebibliography}{50}
\bibitem{nano} E. Roduner, {\it Chem. Soc. Rev.} {\bf 35}, 583 (2006); A.M. Smith and S. Nie, {\it Acc Chem Res.} {\bf 43}, 190 (2010).
\bibitem{hoff}M.R. Hoffman, et al., {\it Chem. Rev.} {\bf 95}, 69 (1995).
\bibitem{maldoti}A. Maldoti, et al., {\it Chem. Rev.} {\bf 102}, 3811 (2002).
\bibitem{bonhole}P. Bonhole, {\it Thin Solid Films} {\bf 350}, 269 (1999).
\bibitem{phillips}L.G. Phillips, et al., {\it J. Dairy Sci.} {\bf 80}, 2726 (1997).
\bibitem{chemmater}D. Dambournet, I. Belharouak and K. Amine {\it Chem.
Mater.} {\bf 22}, 1173 (2009).
\bibitem{luca}V. Luca, et al., {\it J. Phys. Chem. B} {\bf 102}, 10650 (1998).
\bibitem{3}J. Pascual, J. Camassel and H. Mathieu {\it Phys. Rev. B} {\bf 18}, 6842 (1978).
\bibitem{5}H. Tang, F. L{\'e}vy, H. Berger and P.E. Schmid {\it Phys. Rev. B} {\bf 52}, 7771 (1995).
\bibitem{6}A. Mattsson and L. {\"O}sterlund  {\it J. Phys. Chem. C} {\bf 114}, 14121 (2010).
\bibitem{7}U. Bach et al., {\it Nature} {\bf 395}, 583 (1998).
\bibitem{8}M. Gr{\"a}tzel., {\it MRS Bull} {\bf 30}, 23 (2005). 
\bibitem{9}A. Hagfeldt et al., {\it Chem. Rev.} {\bf 110}, 6595 (2010).
\bibitem{tio2review} X. Pan et al., {\it Nanoscale} {\bf 5}, 3601 (2013).
\bibitem{chen}X. Chen and S. S. Mao, {\it Chem. Rev.} {\bf 107}, 2891 (2007). 
\bibitem{60}W. Choi, A. Termin and M. R. Hoffmann, {\it J. Phys. Chem.}
{\bf 98}, 13669 (1994).
\bibitem{64}Z.-R. Tang, Y. Zhang and Y.-J. Xu, {\it ACS Appl. Mater.
Interfaces} {\bf 4}, 1512i (2012).
\bibitem{65}D. Li et al.,
{\it Environ. Sci. Technol.} {\bf 42}, 2130 (2008).
\bibitem{66}Y. Zhang, Z.-R. Tang, X. Fu and Y.-J. Xu, {\it ACS Nano} {\bf 5},
7426 (2011).
\bibitem{70}N. Zhang, Y. Zhang, X. Pan, M.-Q. Yang and Y.-J. Xu, {\it J. Phys.
Chem. C} {\bf 116}, 18023 (2012).
\bibitem{72}Y.-J. Xu, Y. Zhuang and X. Fu, {\it J. Phys. Chem. C} {\bf 114},
2669 (2010).
\bibitem{25}G.-S. Li, D.-Q. Zhang and J. C. Yu, {\it Environ. Sci. Technol.}
{\bf 43}, 7079 (2009).
\bibitem{29}S. Liu, N. Zhang, Z.-R. Tang and Y.-J. Xu, {\it ACS Appl. Mater.
Interfaces} {\bf 4}, 6378 (2012).
\bibitem{77}J. Nowotny, T. Bak, M. K. Nowotny and L. R. Sheppard, {\it Int. J.
Hydrogen Energy} {\bf 32}, 2630 (2007).
\bibitem{79}M. K. Nowotny, L. R. Sheppard, T. Bak and J. Nowotny, {\it J.
Phys. Chem. C} {\bf 112}, 5275 (2008).
\bibitem{80}G. Pacchioni, {\it ChemPhysChem} {\bf 4}, 1041 (2003).
\bibitem{81}I. Nakamura et al.,{\it J. Mol. Catal. A: Chem.} {\bf 161}, 205 (2000).
\bibitem{83}T. Thompson and J. Yates, {\it Top. Catal.} {\bf 35}, 197 (2005).
\bibitem{84}Z. Zhang et al., {\it J. Am. Chem. Soc.} {\bf 128}, 4198 (2006).
\bibitem{gu}T. Gu, {\it J. Appl. Phys.} {\bf 113}, 033707 (2013).
\bibitem{wien2k}P. Blaha, K. Schwarz, G.K.H. Madsen, D. Kvasnicka, and J. Luitz,
{\it WIEN2k, An Augmented Plane Wave + Local Orbitals Program for Calculating
Crystal Properties} (Wien:Karlheinz Schwarz, Techn. Universität Wien, 2001).
\bibitem{dft}T.L. Loucks, {\it Augmented Plane Wave Method}
 (New York:Benjamin, 1967); O.K. Andersen, {\it Solid State Commun.} {\bf 13},
133 (1973); E. Wimmer, et al., {\it Phys. Rev. B} {\bf 24}, 864 (1981).
\bibitem{haule}K. Haule, C. Yee, and K. Kim, {\it Phys. Rev. B} {\bf 81},
195107 (2010).
\bibitem{kim}K. kim {\it {et al.}}, {\it Nat. Mater.} {\bf 17}, 794 (2018).
\bibitem{eug}A. Eugene {\it {et al.}}, {\it Science} {\bf 359}, 186 (2018).
\bibitem{at1}A. Taraphder, S. Koley, N.S. Vidhyadhiraja, and M.S. Laad,
{\it Phys. Rev. Lett.}, {\bf 106}, 236405 (2011).
\bibitem{su}S. Koley, M.S. Laad, N.S. Vidhyadhiraja, and A. Taraphder, {\it Phys. Rev. B}, {\bf 90}, 115146 (2014).
\bibitem{kotliarrmp}A. Georges {\it {et al.}}, {\it Rev. Mod. Phys.} {\bf 68}, 13 (1996).
\bibitem{supc}S. Koley, {\it Solid State Commun.} {\bf 251}, 23 (2017).
\bibitem{aginter}S. Koley, and S. Basu, {\it arXiv:1904.03698}.
\bibitem{haule1} K. Haule, {\it Phys. Rev. B} {\bf 75}, 155113 (2007).
\bibitem{jarrell}M. Jarrell and J. E. Gubernatis, {\it Phys. Rep.} {\bf 269}, 133 (1996).
\bibitem{45}S. Wemple, {\it The Journal of Chemical Physics} {\bf 67}, 2151 (1977).
\bibitem{56}R. Gonzalez, R. Zallen, and H. Berger, {\it Phys. Rev B} {\bf 55}, 7014 (1997).
\bibitem{58}M. O. Atambo et al., {\it Phys. Rev Materials} {\bf 3} 045401 (2019).
\bibitem{arxiv10569}P. Basera, S. Saini, and S. Bhattacharya, {\it arxiv:1905.10569}.
\end{thebibliography}
\end{document}